\documentstyle[11pt,epsfig]{article}
\newcommand{\E}{{\rm E}}

\newcommand{\Cov}{{\rm Cov}}

\newcommand \be  {\begin{equation}}
\newcommand \bea {\begin{eqnarray}}
\newcommand \ee  {\end{equation}}
\newcommand \eea {\end{eqnarray}}

\def \vr {{\vec r}}
\def \ve {{\vec \eta}}
\def \vu {{\vec u}}
\def \vh {{\vec h}}

\def \vp {{\vec p}}
\def \vo {{\vec \omega}}

\def \vmg {{\vec \mu}}
\def \ms {{\mathbf{\Sigma}}}
\def \mO {{\mathbf{O}}}
\def \mOt {{\mathbf{O}^t}}

\def \mQ {{\mathbf{Q}}}
\def \mQt {{\mathbf{Q}^t}}

\def \mL {{\mathbf{\Lambda}}}

\def \mD {{\mathbf{D}}}
\def \mD {{\mathbf{D}}}

\def \vet {{^t{\vec \eta}}}
\def \cN {{\cal N}}

\begin{document}

\title{{\bf Multifractal returns and Hierarchical Portfolio Theory}
\thanks{We are grateful to E.~Bacry, U.~Frisch and L. Martellini for
helpful discussions.}}

\author{\bf J.-F. Muzy$^1$, D. Sornette$^{2,3}$, J. Delour$^1$ and
A.~Arneodo$^1$\\
$^1$ Centre de Recherche Paul Pascal, CNRS UPR 8641\\
Universit\'e de Bordeaux I, Av. Schweitzer, 33600 Pessac, France\\
$^2$ Institute of Geophysics and
Planetary Physics and Department of Earth and Space Science\\
University of California, Los Angeles, California 90095\\
$^3$ Laboratoire de Physique de la Mati\`{e}re Condens\'{e}e\\ CNRS UMR6622 and
Universit\'{e} de Nice-Sophia Antipolis\\ B.P. 71, Parc
Valrose, 06108 Nice Cedex 2, France \\
e-mails: sornette@moho.ess.ucla.edu
}

\date{{\normalsize First Version: April 1999 ~~~~~ This Version: August 2000}}

\maketitle

\begin{abstract}
\noindent

We extend and test empirically the
multifractal model of asset returns based on a multiplicative cascade of
volatilities
from large to small time scales. Inspired by an
analogy between price dynamics and hydrodynamic turbulence
[Ghashghaie {\it et al.}, 1996; Arneodo {\it et al.}, 1998a], it
 models the time scale dependence of the
probability distribution of returns in terms
of a superposition of Gaussian laws, with a log-normal
distribution of the Gaussian variances. This multifractal description
of asset fluctuations is generalized into a multivariate framework
to account simultaneously for correlations across times scales
and between a basket of assets.
The reported empirical results show that this extension is
pertinent for financial modelling. Two sources of departure from normality are
discussed: at large time scales, the distinction between discretely and
continuously
discounted returns lead to the usual log-normal deviation from normality;
at small
time scales, the multiplicative cascade process leads to multifractality
and strong deviations from normality. By perturbation expansions, we are able
to quantify precisely on the cumulants of the distribution of returns,
the interplay and crossover between these two mechanisms.
The second part of the paper applies this theory to portfolio
optimisation. Our multi-scale description allows us to
characterize the portfolio return
distribution at {\it all} time scales {\it simultaneously}. The portfolio
composition is predicted to change with the investment time horizon (i.e.,
the time
scale) in a way
that can be fully determined once an adequate measure of risk is chosen.
We discuss the use of the fourth-order cumulant and of utility functions.
While the portfolio volatility
can be optimized in some cases for all time horizons, the kurtosis and
higher normalized
cumulants cannot be simultaneously optimized.
For a fixed investment horizon, we study in details the influence of
the number of periods, i.e., of the number of rebalancing of the portfolio.
For the large risks quantified by the cumulants of order larger than two,
the number of periods has a non-trivial influence, in contrast with Tobin's
result valid in the mean-variance framework.
This theory provides
a fundamental framework for the conflicting optimization involved in the
different time horizons and quantifies systematically
the trade-offs for an optimal inter-temporal portfolio optimization.

\end{abstract}

\thispagestyle{empty}

\pagenumbering{arabic}

\newpage

\section{Introduction}

Inspired by an analogy with turbulent
cascades in hydrodynamics comparing the energy flow to an information transfer,
recent empirical works
have shown that return volatilities exhibit long-range correlations
organized in a hierarchical way, from large time scales to small time scales
[Ghashghaie {\it et al.}, 1996; Arneodo {\it et al.}, 1998a; Muzy {\it et al.}, 2000;
Breymann {\it et al.}, 2000]. The Olsen group in Zurich discovered independently
this phenomenon which they called the HARCH effect [M\"uller {\it et al.}, 1997;
Dacorogna {\it et al.}, 1998]:
in the foreign exchange market,
the coarse-grained volatility predicts the fine-grained volatility better than
the other way around. They also found this effect for the implied forward rates
derived from Eurofutures contracts [Ballocchi {\it et al.}, 1999].

The underlying cascade or
hierarchical structure provides also a natural explanation and
a model for the multifractal
description of stock market prices documented by several authors
[Fisher {\it et al.}, 1997; Mandelbrot, 1997; Vandewalle and Ausloos, 1998;
Brachet {\it et al.}, 1999;
Bershadskii, 1999; Ivanova and Ausloos, 1999; Mandelbrot, 1999; Schmitt et
al. 1999; Pasquini and Serva, 2000]
(see however Bouchaud {\it et al.} (2000)).

The first purpose of the
present paper is to provide additional empirical confirmations that such
hierarchical description accounts very well for the return
statistics, especially for the strong leptokurticity of the
probability density functions at small scales and for the volatility
serial correlations. This hierarchical framework shares a crucial
advantage with models based on the standard geometrical Brownian motions or on
L\'evy-stable processes, in that
it remains a ``scale-free'' description which can thus be applied to
any time scale.  As we shall see, this ``self-similarity'' property
is interesting for the problem of portfolio optimization, in particular
for the multi-period selection problem.

The portfolio optimization problem consists in finding the optimal
diversification
on a set of possibly dependent assets in order to maximize return and
minimize risk. In its simplest version, one assumes a single period horizon for
all investors which, together with the hypothesis that returns are normally
distributed,
leads to the standard Markowitz's solution [Levy and Markowitz, 1979; Kroll
{\it et al.}, 1984].
The first results in the theory of optimal multi-period portfolio
selection showed that, within the Gaussian hypothesis of return distributions,
the investor optimal sequence of portfolio through time is stationary, with
constant proportionate holdings of each included asset leading to a constant
expected return and risk per unit invested wealth. In other words, the optimal
asset weights in the portfolio are time-scale independent. This theorem has
been
criticised as being not true in general [Stevens, 1972].

Many studies have investigated
the impact of time horizon on the portfolio selection. Allowing for
investors with different planning horizons,  Gressis {\it et al.} [1976] have shown
that, in absence of the riskless security into the portfolio, the single-period
efficient frontier also provides $K$ periods efficient frontiers. However,
Gressis {\it et al.} [1976] find that, with the inclusion of the riskless security,
the equivalence among efficient frontiers for different horizons is no
longer valid.
Departing from the Gaussian model,
Arditti and Levy [1975] generalized the mean-variance approach to a
three-moment efficiency analysis and showed that all one-period portfolios
are not necessarily multi-period efficient because, even if a stock has a
symmetric
single-period return distribution, its multi-period distribution may be
highly skewed.
A considerable body of research has explored how portfolio
 composition depends on the investment horizon. It has been shown that
portfolio composition indeed depends on the investment horizon and that
any simple characterization of the relationship is treacherous [Gunthorpe
and Levy, 1994;
Ferguson and Simaan, 1996].
Using a choice criterion that is
consistent with the traditional utility approach but which is more
amenable to a multiperiod environment, Marshall [1994] has shown that
investors should choose progressively less risky single-period portfolios
as their
investment horizons shorten, even if they do not become more risk averse,
both in the presence and in the absence of a riskless asset.
Tang [1995] examined the effect of investment horizon on international
 portfolio diversification using stock indexes of 11 countries.  He finds that
correlation coefficients between stock indexes
increase in general with an increase in the investment horizon, suggesting
that diversification benefits are reduced.  The fact that various stock
markets are more correlated over a longer investment horizon may imply
that different stock markets adjust to each other with a delayed pattern.
Bierman [1997; 1998] has investigated whether the risk of a stock portfolio
increases or
decreases as the investment horizon lengthens and whether the portfolio mix
depends on the horizon.
By moving beyond the logarithmic utility function and by recognizing that
different investors have different risk preferences, Bierman [1998]
finds that it is possible for stocks to be risky in each
time period and to reduce risk by increasing the length of the investment
horizon. He also finds that stocks can be rejected if they are to be held
for one time period, but can then be accepted if they are to be held for more
than one time period. At the origin of these results is the fact that
``discrete'' returns $r$ calculated as the ratio of the price difference
over a given time interval
$\tau$ over the price are only approximately proportional to $\tau$ times the
continuously discounted return $\eta$. The difference is negligible at
short times $\tau$,
corresponding to the validity
of the expansion of $e^{\eta \tau} = 1 + \eta \tau$. At large time scales
such that
$\eta \tau$ is no more small compared to $1$, the second-order
term in the expansion becomes important and effets of non-normality in the
distribution of
$r$ become important. In the present paper, we study another additional
cause for
non-normality, whose strength grows as short time scales rather than at
large time scales.

The paper is organized as follows. In section 2, we recall and make precise
the statistical description of the fluctuations of the returns of a
single asset in terms of the multiplicative cascade model.
The few parameters involved in this model are then interpreted and
estimated for a set of high frequency time series.
Then, we propose an extension of this framework to
vector valued, i.e. multivariate processes corresponding to a basket of
assets in
a portfolio.
This can be done in a very natural way and simple statistical
tests are proposed to check for its relevance.
In section 3, we address the portfolio problem for such a set
of multifractal assets. We first discuss the simplest case of uncorrelated
assets where
the statistical properties of the portfolio returns can be estimated
by means of a cumulant expansion.
The problem of portfolio optimization is studied in section 4 using
the utility function approach and the higher-order cumulants of the
distribution of
portfolio returns. A generalisation of the efficient frontier is proposed and
studied as a function of time-scales.
Section 5 presents the application of the correlated
multivariate multifractal model to the portfolio theory.
Conclusion and prospects are given in section 6.

\section{Multifractal description of asset returns}

In the following, we show how multifractal statistics can be
described
in a simple way using only a few parameters. This description
has been developed both for finance data [Arneodo {\it et al.}, 1998a]
and in the field of fully-developed hydrodynamic turbulence
[Arneodo {\it et al.}, 1998b; 1998c, 1999].
As mentionned above, the similarity between turbulence and finance has been
suggested by some recent studies [Ghashghaie {\it et al.}, 1996; Arneodo {\it et al.},
1998a; Muzy {\it et al.}, 2000]
and relies on the concept
of multiplicative random cascades: the fluctuation at some
fine time scale $\tau_1$ is the product of the fluctuation at a
coarser scale $\tau_2$ by a random factor whose law depends
only on the scale ratio $\tau_2/\tau_1$. This picture leads to
the integral description proposed by B. Castaing and collaborators
[Castaing {\it et al.}, 1990; Chabaud {\it et al.}, 1994]
which links the probability density function (pdf) of fluctuations at scale
$\tau_1$
to the pdf of fluctuations at scale $\tau_2$. Let us now introduce this
formalism for the pdf's of asset returns for all time scales.

\subsection{Formulation of the multifractal model of the returns of a
single asset}

Let $p_i(t)$ be the price of asset $i$ at time $t$, where time is counted
for trading days in
multiples of a fundamental unit (say days). With the notation $\eta_i
(t,\tau) =
\ln\left({p_i(t) \over p_i(t-\tau)}\right)$, the return
$r_i(t,\tau)$ between time $t-\tau$ and $t$ of asset $i$ is defined as:
\be
 r_i(t,\tau) = {p_i(t)-p_i(t-\tau) \over p_i(t-\tau)} = e^{\eta_i (t,\tau)}
-1, \label{jhghss}
\ee
We thus distinguish between the ``continuous return'' $\eta_i (t,\tau)$
and the ``discrete'' return $r_i(t,\tau)$. The difference between them
is essentially captured by the
second order correction ${1 \over 2}[\eta_i (t,\tau)]^2$,
which becomes non-negligible only for large time scales (see below).

We describe the distribution of the variable $\eta_{i}(t,\tau)$ by using
the representation of Castaing {\em et al.} [1990; Chabaud {\it et al.}, 1994]
inspired by the above mentionned cascade picture for hydrodynamic
turbulence. This
distribution is represented, for time scale $\tau_1$, as a
weighted sum of
dilated distributions of the $\eta$'s
at the intermediate coarser scale
$\tau_2$ ($\tau_1 \leq \tau_2$):
\be
P_{i,\tau_1}(\eta_{i} - \mu_i \tau_1) = \int_{-\infty}^{+\infty} du \;
e^{-u} ~G_{i,\tau_2 \rightarrow \tau_1}(u) ~P_{i,\tau_2} (e^{-u} (\eta_{i}
-\mu_i
\tau_2))~,
\label{jnanhak}
\ee
where $\mu_i \tau$ is the mean value of $\eta_{i}$ at scale $\tau$
(it is supposed to be linear with $\tau$ consistently with experimental
observations)
and $G_{i,\tau_2 \rightarrow \tau_1}(u)$ is the {\em self-similarity kernel}
whose shape is assumed to depend only on the ratio $\tau_2/\tau_1$. As we
shall see below,
the exponentials $e^{-u}$ are just convenient ways of simplifying the
parameterization of the kernel.
This model (\ref{jnanhak}) can be derived from a cascade model
according to which $\eta(t,\tau_1)$ is written as
\be
\eta(t,\tau_1) = \sigma(t,\tau_1) X(t)~,
\ee
 where $X(t)$ is a $\tau_1$-independent
Gaussian random variable and the ``stochastic volatility'' $\sigma(t,\tau_1)$
is described by a multiplicative cascade [Arneodo {\it et al.}, 1998a]:
\be
\sigma(t,\tau_1)= W_{\tau_2 \rightarrow \tau_1} \sigma_{t,\tau_2}~.
\label{nvbbx}
\ee

It is a well-established observation
that, at sufficiently large scale,
distributions of $\eta$ become Gaussian [Campbell {\it et al.}, 1997].
Let $T_i$
be such a large scale for which $P_{i,T_i} (\eta_{i} - \mu_i T_i)$ is a
Gaussian $\cN(0,\sigma^2_{T_i})$. Then, for $\tau \leq T_i$, expression
(\ref{jnanhak}) can be rewritten as
\be
P_{i,\tau}(\eta_{i} - \mu_i \tau) = \int_{-\infty}^{+\infty} du \; ~G_{i,T_i
\rightarrow \tau}(u) ~\cN_{i,T_i}(\eta_{i} - \mu_i T_i,e^{2u}
\sigma_{T_{i}}^{2})~.
\label{aear}
\ee
$P_{i,\tau}(\eta - \mu_i \tau)$ is thus a weighted sum of Gaussian
distributions and the kernel $G_{i,T_i \rightarrow \tau}(u)$ can be
identified with
the pdf of the logarithm of the multiplicative weights
$u \equiv \ln(W_{T_i\rightarrow \tau})$ introduced in Eq.(\ref{nvbbx}),
the standard deviation at scale $\tau$ being:
\be
\sigma_i= e^u \sigma_{T_i} ~.  \label{fjnball}
\ee
Let $h_i(\tau)$ and $\lambda_i^2(\tau)$ be respectively the mean value and
the variance of $G_{i,T_i\rightarrow \tau}$. We consider the simplest
possible case,
specifically where $G_{i,T_i \rightarrow \tau}(u)$ is itself Gaussian:
\be
G_{i,T_i \rightarrow \tau}(u) = {1 \over \sqrt{2\pi} ~\lambda_i(\tau)}~\exp
\left( -{\left(u-h_i(\tau)\right)^2 \over 2 \lambda_i(\tau)^2} \right)~.
\label{qkmqmmq}
\ee
From the semi-group property resulting from the cascade equation,
$W_{\tau_3\rightarrow \tau_1}= W_{\tau_3\rightarrow
\tau_2}W_{\tau_2\rightarrow \tau_1}$,
and using the fact that $G_{i,\tau_2\rightarrow \tau_1}$ depends only on
the ratio
$\tau_2/\tau_1$, it is easy to show that $h_i(\tau)$ and
$\lambda_i^2(\tau)$ are
necessarily
linear functions of $\ln(\tau)$:
\bea
h_i(\tau) &=& a_i + h_i \ln \tau~, \label{ghljkqq}\\
\left[\lambda_i(\tau)\right]^2 &=& b_i - \lambda_i^2 \ln \tau~, \label{lqdqm}
\eea
where $a_i, b_i, \lambda_i$ and $h_i$ are independent of $\tau$.
$T_i$ is the coarse scale defined by a vanishing variance
$\lambda_i^2(T_{i}) = 0$. This time scale $T_i$ is the ``integral'' scale at
which the cascading
process of the volatility begins. Performing the change of variable
$u \to u - h_i(T_i)$ amounts to keep Eq.(\ref{aear})
unchanged and to redefine
Eqs (\ref{ghljkqq}) and (\ref{lqdqm}) as :
\bea
h_i(\tau) &=& h_i \ln (\tau/T_i) ~, \label{hi}\\
\left[\lambda_i(\tau)\right]^2 &=& -\lambda_i^2 \ln (\tau/T_i)~. \label{li}
\eea
As shown in the next section, these equations account
very well for empirical observations.

Let us note that this description (\ref{aear}) with (\ref{qkmqmmq}) and
(\ref{li}) contains
the standard Gaussian random walk model
of stock market returns as the special case :
\bea
h_i &=& {1 \over 2}~, \\
\lambda_i &=& 0~.
\eea
Indeed, for $\lambda_i \to 0$, $G_{i,T_i \rightarrow \tau}(u)$ becomes
a delta-function centered on $u=h_i$, i.e. the volatility (\ref{fjnball})
at scale $\tau$ is $\sigma_i= \sigma_{T_i} e^u = \sigma_{T_i} e^{h_i}
= \sigma_{T_i} \left({\tau \over T_i}\right)^{1/2}$, which recovers
the standard square-root diffusion law. In this case,
the pdf of asset ``continuous'' returns
$P_{i,\tau}(\eta_{i} - \mu_i \tau)$ given by Eq.(\ref{aear})
is Gaussian.

\subsection{Empirical tests}

In order to test the previous formalism empirically, we need to
define estimators for the parameters $T_i$, $\sigma_{T_i}^2$,
$h_i$ and $\lambda_i^2$. For this purpose, the
so-called ``structure functions'' or $q$-order moments $\langle
\eta^q_i(t,\tau) \rangle$
of the ``continuous'' returns are very useful quantities
(the brackets define the average with respect to a running window along the
time series). For the purpose of analyzing data with multiple
scales, the dependence of $\langle \eta^q_i(t,\tau)
\rangle$ as a function of time scale $\tau$ for various $q$'s is very
rich in
information [Frisch, 1995].
For the hierarchical model (\ref{aear}) with the Gaussian
ansatz given by Eq.(\ref{qkmqmmq}),
the $q$th-order structure function of the centered variable
$\eta'_i(t,\tau) = \eta_i(t,\tau)-\mu_i \tau$
(whose law is denoted using prime)
\be
M_i(q,\tau) = \langle {\eta'}_i^q \rangle = \int_{-\infty}^{+\infty}
\eta'^q_{i}
P'_{i,\tau}(\eta'_{i})
d\eta'_{i} \;,
\ee
can be calculated analytically with the change of variable
$R = \eta'_{i} e^{-u}$:
\be
M_i(q,\tau) = \left(\int_{-\infty}^{+\infty} e^{qu}
G_{i,T_i
\rightarrow \tau}(u) du \right) \left(\int_{-\infty}^{+\infty} R^q
P'_{i,T_i} (R) dR \right)~.
\label{hfbva}
\ee
The second factor in the r.h.s. of Eq.(\ref{hfbva})
is the $q$th-moment of the distribution at the integral scale.
All the dependance on the time scale $\tau$ is found in the first factor
that can be computed using Eqs (\ref{qkmqmmq}),(\ref{hi}) and (\ref{li}):
\be
\int_{-\infty}^{+\infty} e^{qu} G_{i,T_i \rightarrow \tau}(u) du =
\left({\tau \over T_i}\right)^{\zeta_i(q)} ~,  \label{jfaagjhajak}
\ee
with
\be
\label{zetaln}
\zeta_i (q) = h_i q - {\lambda^2_i \over 2} q^2~. \label{jnbvgaknva}
\ee
We thus get
\be
M_i(q,\tau) = \left(\int_{-\infty}^{+\infty} R^q
P'_{i,T_i} (R) dR \right)~~\left({\tau \over T_i}\right)^{\zeta_i(q)}~.
\label{ngwerw}
\ee
The dependence of the $q$th-order moments
of ``continuous'' returns as a function of time scale is a pure power law,
resulting from
the logarithmic behavior as a function of time scale
of both the mean and the variance of the
Gaussian self-similarity kernel (Eqs. (\ref{hi}) and (\ref{li})).
The exponent $\zeta_{i}(q)$ is related to the cumulant generating function of
the kernel, as shown by Eq.(\ref{jfaagjhajak}). In general, the centered
moments of
odd-orders at the integral scale are vanishing due to the approximate symmetric
structure of the pdf's. In particular, the first-order moment, the centered
``continuous'' return, is zero by definition. From expression
(\ref{ngwerw}), the same property
follows for
$\langle {\eta'}_i^q \rangle$.
In order to measure the power law dependence given by the second
term $(\tau / T_i)^{\zeta(q)}$ of the r.h.s. of Eq.(\ref{ngwerw}),
it is convenient to
take the absolute value of the ``continuous'' return and calculate :
\be
\langle |{\eta'}_i|^q \rangle
\propto \left({\tau \over T_i}\right)^{\zeta_i(q)}~.  \label{mjvfnbzz}
\ee

In order to determine the
two parameters $h_i$ and $\lambda_i$, we measure the
``multifractal spectrum'' $\zeta_i(q)$ by performing a linear fit of
$\ln M_i(q,\tau)$ as a function of $\ln \tau$,
for a wide range of values $q$. The slope of the linear fit gives the exponent
$\zeta_i(q)$. The slope at the origin of $\zeta_i(q)$ as a function of $q$
provides an estimation of the parameter $h_i$.  The fit of $\zeta_i(q)-h_i q$
then provides a check for the importance of the quadratic
correction proportional to $\lambda_i^2$. Recall that, for a Gaussian pdf
of ``continuous'' returns,
$\lambda_i^2 =0$ and the exponents $\zeta_i(q)$ are linear in $q$. The
quadratic
dependence, often called ``multifractal'' in the literature [Frisch, 1995], is
a signature of the multi-scale structure.

\begin{figure}
\begin{center}
\epsfig{file=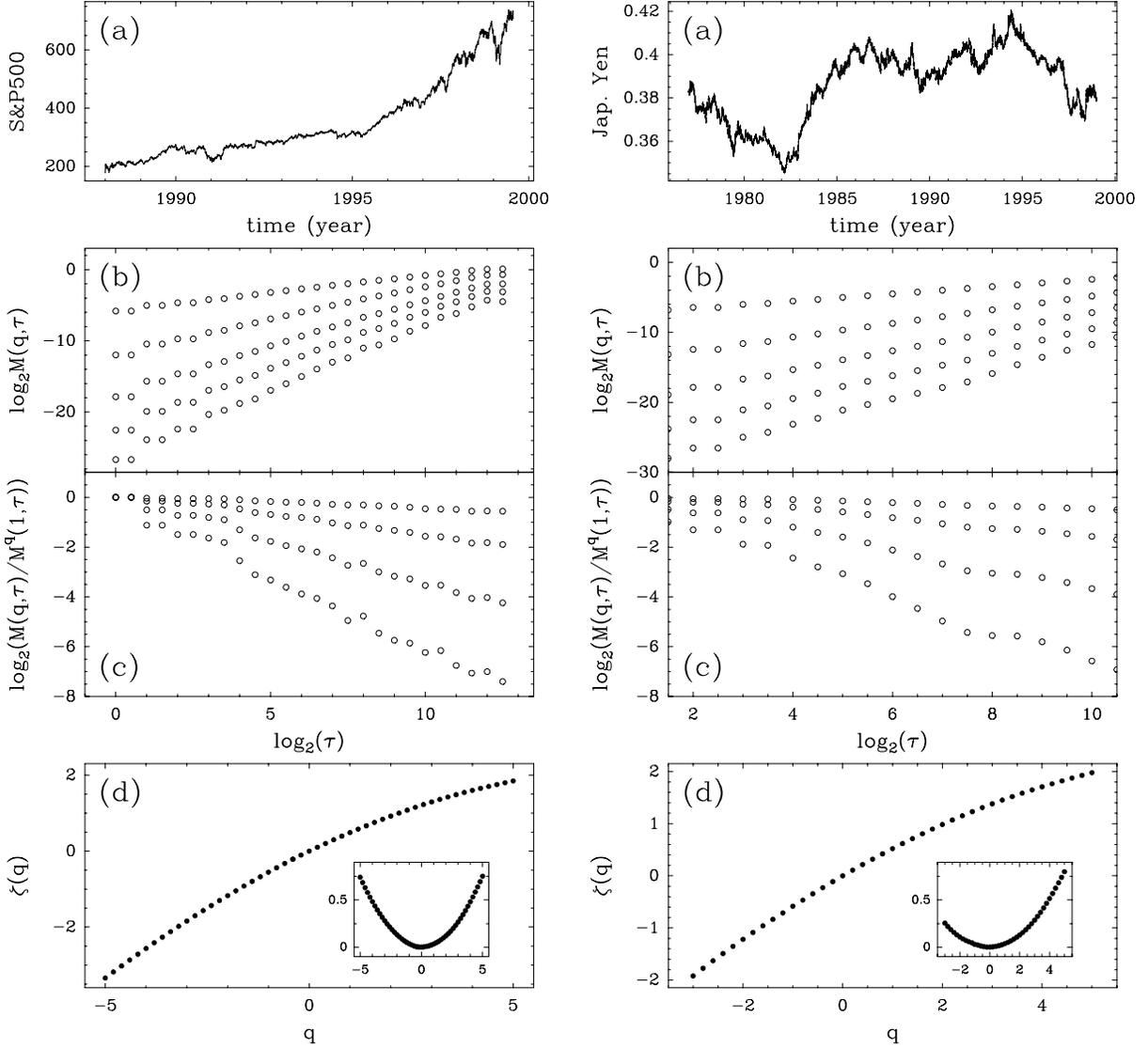, width=16cm}
\caption{\protect\label{fig1} (a) the Standard \& Poor US index
(S\&P500) time series (left panel) and the Japanese Yen time series (right
panel)
with a sampling time of ten minutes are shown as a function of time
over more than one (resp. two) decades. Intraday ``seasonal''
effects (large volatility at the open and close) are removed before calculating
the moments. (b) $\log_2(M(q,\tau))$ versus $\log_2(\tau)$ for $q=1,2,3,4,5$.
The slope of these curves provides an estimate of $\zeta (q)$. (c)
$\log_2(M(q,\tau)/M^q(1,\tau))$ versus $\log_2(\tau)$ for $q=2,3,4,5$ (top
to bottom); deviation
from a constant is the signature of ``multifractality''.
(d) $\zeta(q)$ spectra for the two analyzed series. As illustrated in
the insets where the linear part of the spectra has been
removed, they are almost perfectly fitted by a pure
parabola with quadratic coefficient $\lambda^2 = 0.018$ (S\&P500) and
$\lambda^2 = 0.032$
(Japanese Yen).
We find $h - \lambda^2 = 1/2$ to a very good approximation (see text for
an explanation).
}
\end{center}
\end{figure}

In Fig.~\ref{fig1}, we report such an analysis for both
the S\&P500 index future and Japanese yen/US dollar exchange
rate time series shown at the top in the two panels (a).
The original intraday series have been sampled at a 10 mn rate
and seasonal effects on the volatility have been removed. One can
see in Figs~\ref{fig1}(b) that the moments $M_{i}(q,\tau)$
do exhibit the predicted scaling (Eq. (\ref{mjvfnbzz})) over a large range of
scales, of which three decades are here shown.
The multifractal
nature of these series is illustrated in Figs~\ref{fig1}(c) which shows
the ratio ${M_{i}(q,\tau) \over M_{i}(1,\tau)^q}$ for
$q=2, 3, 4$ and $5$ in log-log coordinates. For linear $\zeta_{i}(q)$
functions,
the ratios ${M_{i}(q,\tau) \over M_{i}(1,\tau)^q}$ should be constants as a
function of time scales.
A departure from a constant thus qualifies a multifractal behavior. The
dependence of the
exponent $\zeta_{i}(q)$ reported in Figs~\ref{fig1}(d) is very well fitted by
the parabolic shape given by Eq. (\ref{zetaln}).
We find $h_{i} - \lambda_{i}^2 = 1/2$ to a very good approximation for all
assets which, according to Eq. (\ref{jnbvgaknva}),
ensures that the volatility or variance $\langle {\eta'}_i^2 \rangle \sim t$ as
for the standard geometrical Brownian model,
i.e. does not exhibit an anomalous scaling.

\begin{figure}
\begin{center}
\epsfig{file=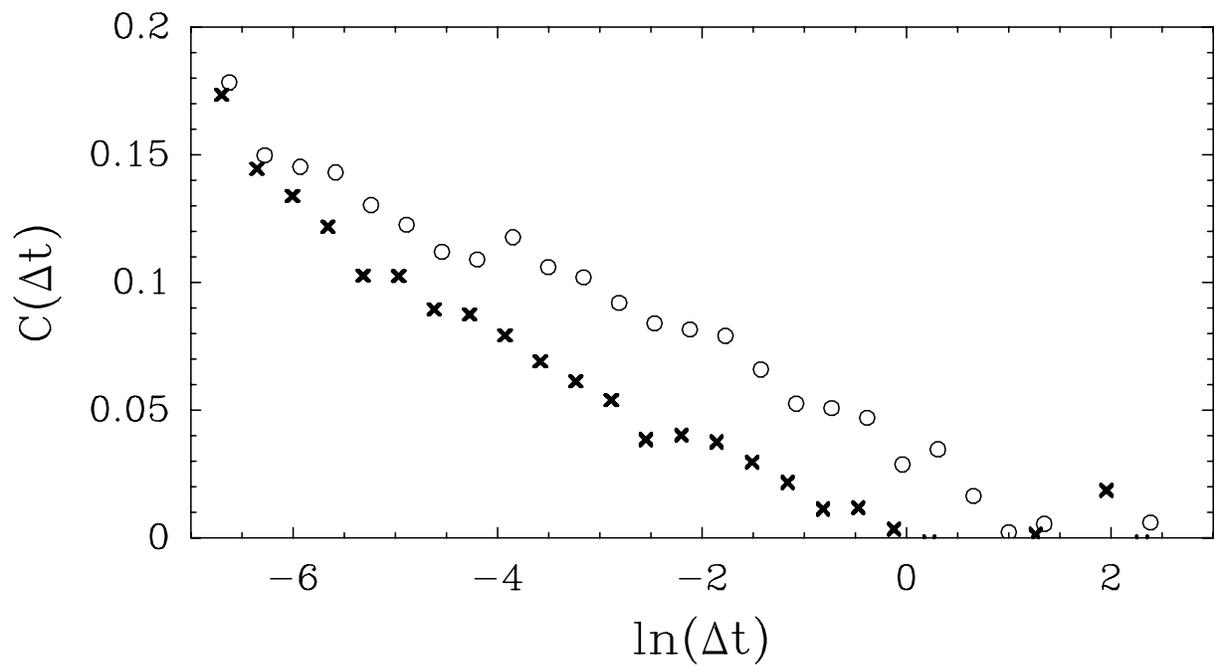, width=16cm}
\caption{\protect\label{fig2} This figure shows the autocorrelation
function defined by
Eq. (\ref{mfmzsq}) of the
``magnitudes'' $\omega_i(t,\tau) = \ln(|\eta_i(t,\tau)|)$ for $\tau = 10$
minutes for the SP\&500 ($\circ$) and Japanese Yen ($\times$). A linear regression of these functions
versus the logarithm of time lag $\Delta t$ enables us to evaluate
$\lambda_i^2$
and $T_i$ for each time series.
}
\end{center}
\end{figure}

In order to test for the existence of a cascading process
and to estimate the integral time $T_i$, we refer to
[Arneodo {\it et al.}, 1998a] where it is shown that,
for a cascade process, the covariance of the logarithms of
the $\eta_{i}$'s at all scales should decrease as a logarithmic function.
More precisely, if one defines the {\em magnitude} of asset $i$
at scale $\tau$ and time $t$ as $\omega_i(t,\tau) = \ln(|\eta_i(t,\tau)|)$,
then one should have, for $\Delta t > \tau$:
\begin{equation}
   C_i(\Delta t,\tau)= \Cov(\omega_i(t+\Delta t,\tau),\omega_i(t,\tau)) \simeq
   -\lambda_i^2 \ln({\Delta t \over T_i}) \; .  \label{mfmzsq}
\end{equation}
This behavior is checked for the S\&P500 and
JPY/USD series in Fig.~\ref{fig2} where we have plotted $C(\Delta t, \tau)$
versus $\ln(\Delta t)$ with $\tau=10$mn.
The linearity of these plots is reasonably well verified. The
values of $\lambda_i^2$ and $\ln(T_i)$ can be obtained
respectively from the slope and the intercept of these straight lines.
In Table 1, we have reported the estimates of the parameters
$\lambda_{i}^2$, $\sigma_{T_i}^2$, $\mu_{i}$ and $T_i$ for a set of high
frequency future time series and the mean values of those parameters for
daily time series of stocks composing the Dow Jones Industrial Average index
and the french CAC40 index. Let us note that the errors on the estimates
of $T_i$ can be very large since we get only one estimator of $\ln(T_i)$.
For all series, we have checked that the relationship $h_i-\lambda_i^2=1/2$
is reasonable so we have not reported the estimated values of $h_i$.
We can remark that the values of $\lambda_i^2$ are relatively close to
$0.02$ and the values of $T_i$ are in the range $1-2$ years.
These values can thus be considered as representative
of market multifractality.

\begin{table}
\begin{center}
 \begin{tabular}{|c|c|c|c|c|}
  \hline Series & $\lambda^2$ & $T$ (year) & $\mu_{i}$ (year$^{-1}$) &
$\sigma_{T_i}^2$ (year$^{-1}$)\\
  \hline German Government Bond (F)& 0.028 & 1.2 & -1.6 $10^{-3}$ & 1.7
$10^{-5}$ \\
  \hline FT-SE 100 Index (F)& 0.017 & 1.1 & -3.3 $10^{-3}$ & 7.6 $10^{-5}$ \\
  \hline Japanese Yen (F)& 0.032  & 1.0  & -1.1 $10^{-3}$ & 3.8 $10^{-5}$ \\
  \hline S\&P 500 (F)& 0.018 & 4.5 & 5.7 $10^{-3}$ & 1.8 $10^{-4}$ \\
  \hline Japanese Government Bond (F)& 0.121 & 1.3 & 9.2 $10^{-4}$ & 4.0
$10^{-6}$ \\
  \hline Nikkei 225 (F)& 0.030 & 1.5 & 5.5 $10^{-3}$ & 3.9 $10^{-5}$ \\
  \hline French CAC40 (mean) (S)& 0.020 & 1.9 & 1.2 $10^{-1}$ & 5.6 $10^{-2}$ \\
  \hline Dow-Jones (mean) (S)& 0.010 & 1.7 & 1.7 $10^{-1}$ & 4.6 $10^{-2}$ \\
  \hline
\end{tabular}
 \caption{Estimates of the multifractal parameters for intraday (S\&P 500
and Japanese Yen futures)
 and daily time series. The values measured at different time scales are
consistent with each
 other.}
 \label{tab1}
\end{center}
\end{table}

\subsection{The multivariate multifractal model}

Up to now, our emphasis has been on the correlation structure
within each asset separately.
We have not investigated the impact of correlations across
assets and have quantified only the ``diagonal'' multifractal features as
it is the case for
uncorrelated assets.
Since one of the natural applications of the
characterization of
distributions of returns is the quantification and selection of portfolios
made of
a basket of assets, it is important to generalize our framework to account
for the
possible existence of inter-asset correlations, in addition to the
multi-scale time
correlations. As it is well-known, optimizing a portfolio relies on two
effects, the law
of large numbers and the possibility to counter-act the negative effect of
one asset
by using (anti-) correlations. Towards a practical implementation of our
model, it is thus
essential to offer a generalization of the multifractal cascade model that
accounts for
the correlations across assets.

When dealing with Gaussian
processes, the lack of independence is entirely embodied in the
correlation function. Along this line of thought, we now propose
a phenomenological model that extends the previous
one to the case of correlated assets. We would like
to generalize Eqs.~(\ref{aear}), (\ref{qkmqmmq}),
(\ref{ghljkqq}) and (\ref{lqdqm}) to multivariate distributions.
In this purpose, let us write $\vet$ the line vector
$(\eta_1,...,\eta_N)$,
where the subscript $^t$ stands for `transpose',
and we suppose that there exists a time scale $T$ for which the multivariate
distribution $\cN_{N,T}(\ve,\ms_{0})$ is Gaussian with covariance matrix
$\ms_{0}$. The multivariate Gaussian mixture that generalizes
Eq.(\ref{aear}) can be written as
\be
\label{casmul}
 P_\tau(\ve-\tau \vmg) = \int d^Nu \; G_\tau(\vu) \;
\cN_{N,T}\left(\ve-T \vmg,\mD(\vu) \ms_0 \mD(\vu)\right) ,
\ee
where $\mD(\vu)$ is the diagonal dilation
matrix $\mD_{ij} = e^{u_i} \delta_{ij}$ and $G_{\tau}(\vu)$ is some
multivariate probability distribution function. This description
corresponds to a multivariate cascading process where the
multiplicative weights $W_{\tau_i \rightarrow \tau_j}$ associated with
different assets can be correlated. This form (\ref{casmul}) is the
multivariate generalization of (\ref{aear}).

As in the previous section, let us make a Gaussian ansatz for the joint law
$G$ of the logarithm of these weights, i.e.,
\begin{equation}
   G_{\tau} (\vu) = \cN_N(\vu-\vh(\tau),\mL(\tau)) \; ,
\end{equation}
where the mean vector
$\vh(\tau)$ and the covariance matrix $\mL(\tau)$ are supposed to
behave as:
\bea
   \vh_{i}(\tau) &=& a_i+h_i \ln(\tau) \; , \label{tt1} \\
   \mL_{ij}(\tau) &=& - \lambda^2_{ij} \ln(\tau / T_{ij}) \; , \label{tt2}
\eea
with the ``integral times'' $T_{ij}$ defined by
\be
   \mL_{ij}(\tau=T_{ij}) = 0 \;.
\ee
Within the cascade model, the times $T_{ij}$ are the time scales at
which some cascading process begins. If one considers two assets $i$ and
$j$, it seems natural to assume that only two situations
occur:
\begin{itemize}
  \item[i/] the cascade on $i$ and $j$ has the same ``economic''
origin, and
  then $T_{ii} = T_{jj} = T_{ij}$;
  \item[ii/] the cascades on $i$ and $j$ are independent and thus
   $\mL_{ij}(\tau) = 0$, but, a priori, $T_{ii} \neq T_{jj}$.
\end{itemize}

The $N$ returns $\eta_i$ can be thus organized
by ``classes'' in such a way that the matrix
$\mL$ is a block diagonal matrix:
\be
          \mL = \bigoplus_{l=1}^L \mL_l ~, \label{bd}
\ee
where $L$ is the number of indenpendent subspaces (cascades) of dimension
$N_l$
($\sum_{l=1}^L N_l = N$).
At fixed $l$, one can always redefine $\ms_0$ such that Eqs.~(\ref{tt1})
and (\ref{tt2})
become :
\bea
   \vh_{i}(\tau) &=& h_i \ln(\tau/T_l) \label{nh} \label{hi2} \\
   \mL_{ij}(\tau) &=& - \lambda^2_{ij} \ln(\tau / T_l) \; ~~~~\mbox{for}
\label{lij}\;
 i,j \; \mbox{in the bloc} \; l~. \label{ns}
\eea

To demonstrate that this multifractal multivariate model is compatible with
the empirical observations, we first define an
estimator of the elements of the covariance matrix $\mL$.
Let us denote $\omega_i(t,\tau)$ the magnitude, i.e.,
the logarithm of the ``continuous'' return $\eta$ for asset $i$ at
scale $\tau$: $\omega_i(t,\tau) = \ln(|\eta_i(t,\tau)|)$.
In the multivariate framework,
the structure function at scale $\tau$ can be extended
to order $\vp$, where $\vp$ is a vector of $N$ values, in the
following way:
\begin{equation}
  \langle |\eta_1(t,\tau)|^{p_1}\ldots |\eta_N(t,\tau)|^{p_N}\rangle =
\langle e^{\vo(t,\tau).\vp} \rangle~,
\end{equation}
where $\vo(t,\tau)$ is the vector constructed from
the $\omega_i$'s at scale $\tau$. Using Eq. (\ref{casmul}), the
behavior of the structure function as a function of $\tau$ can be estimated:
\[
 \langle e^{\vo(t,\tau).\vp} \rangle =
 \langle e^{\vo(t,T).\vp} \rangle \int d^Nu \; e^{\vp.\vu} G_\tau(\vu) ~,
\]
where $G$ is the kernel quantifying the cascade from the integral scales to
scale $\tau$.
Using our assumption that $G$ is normal with mean $\vh(\tau)$ and
covariance $\mL(\tau)$,
the characteristic function of $G$, defined as the last term
in the previous expression, is
\[
\int d^Nu \; e^{\vp.\vu} G_\tau(\vu) = e^{\vp.\vh(\tau)}
~e^{\frac{1}{2}^{t}\vp.\mL(\tau).\vp}~.
\]
From the logarithmic behavior in Eqs (\ref{tt1}) and (\ref{tt2}), one deduces:
\begin{equation}
  \langle e^{\vo(t,\tau).\vp} \rangle \sim K_{\vp}~ \tau^{\zeta(\vp)} \; ,
\label{hhgggd}
\end{equation}
with
\begin{equation}
  \zeta(\vp) = \vh.\vp + \frac{1}{2} ^t\vp \mL \vp = \sum_i
p_ih_i-\frac{1}{2}\sum_{i,j}
  p_ip_j\lambda^2_{ij} ~.
\end{equation}
The constant $K_{\vp}$ depends on the values of the integral
scales $T_{ij}$, $h_{i}$, $\lambda_{ij}$ and on the multivariate distribution
at the large integral scales.
This power law (\ref{hhgggd}) can be explicitely used to estimate the
values $\lambda^2_{ij}$
of the elements of the covariance matrix defined by the kernel $G$, by using
different set of $p$'s.

If one sets $p_k=0$ for $k \neq i,j$,
and $p_i=p_j=\epsilon \rightarrow 0$ in the expression (\ref{hhgggd}),
one can show that the covariance between the magnitudes
$\omega_i(t,\tau)$ and $\omega_j(t,\tau)$ simply behaves as:
\begin{equation}
  \Cov(\omega_i(t,\tau),\omega_j(t,\tau)) \simeq -\lambda^2_{ij} \ln(\tau) +
  K_{ij}\; ,
\end{equation}
where $K_{ij}$ depends on $T_{ij}$, $h_{i}$ and $\lambda_{ij}$.

Alternatively, one can test the multivariate
cascade ansatz from the behavior of the lagged magnitude covariance:
in full analogy with the monovariate case, it should behave as:
\begin{equation}
 C_{ij}(\Delta t) = \Cov(\omega_i(t+\Delta t,\tau) \omega_j(t,\tau)) =
 -\lambda^2_{ij}\ln(\Delta t/T_{ij}) \; ,  \label{vnkkd}
\end{equation}
for $\Delta t \geq \tau$.
This equation provides other simple estimators of both $\lambda_{ij}$
and $T_{ij}$.

\begin{figure}
\begin{center}
\epsfig{file=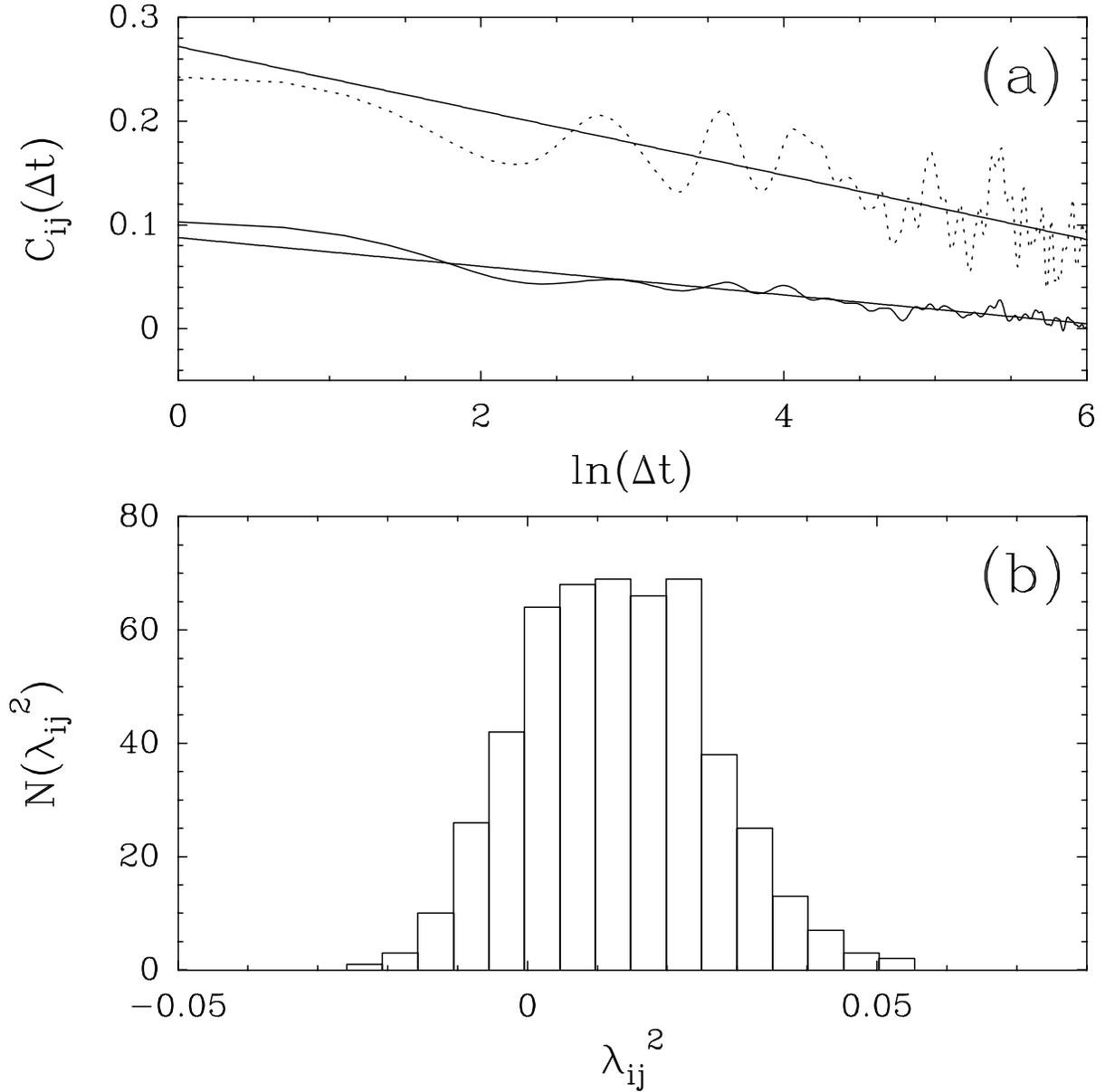, width=16cm}
\caption{\protect\label{m1} Panel (a) represents in dotted line
the magnitude correlation function for the
pair of stocks CCF/MICHELIN of the French CAC40 index. The mean of such
correlation functions over all stock pairs of the CAC40 index is the
continuous line.
The slope and the intercept of these functions allow us to
evaluate $\lambda_{ij}^{2}$ and $T_{ij}$. Panel (b) shows the
histogram of $\lambda_{ij}^{2}$ for all index pairs.
}
\end{center}
\end{figure}

\begin{figure}
\begin{center}
\epsfig{file=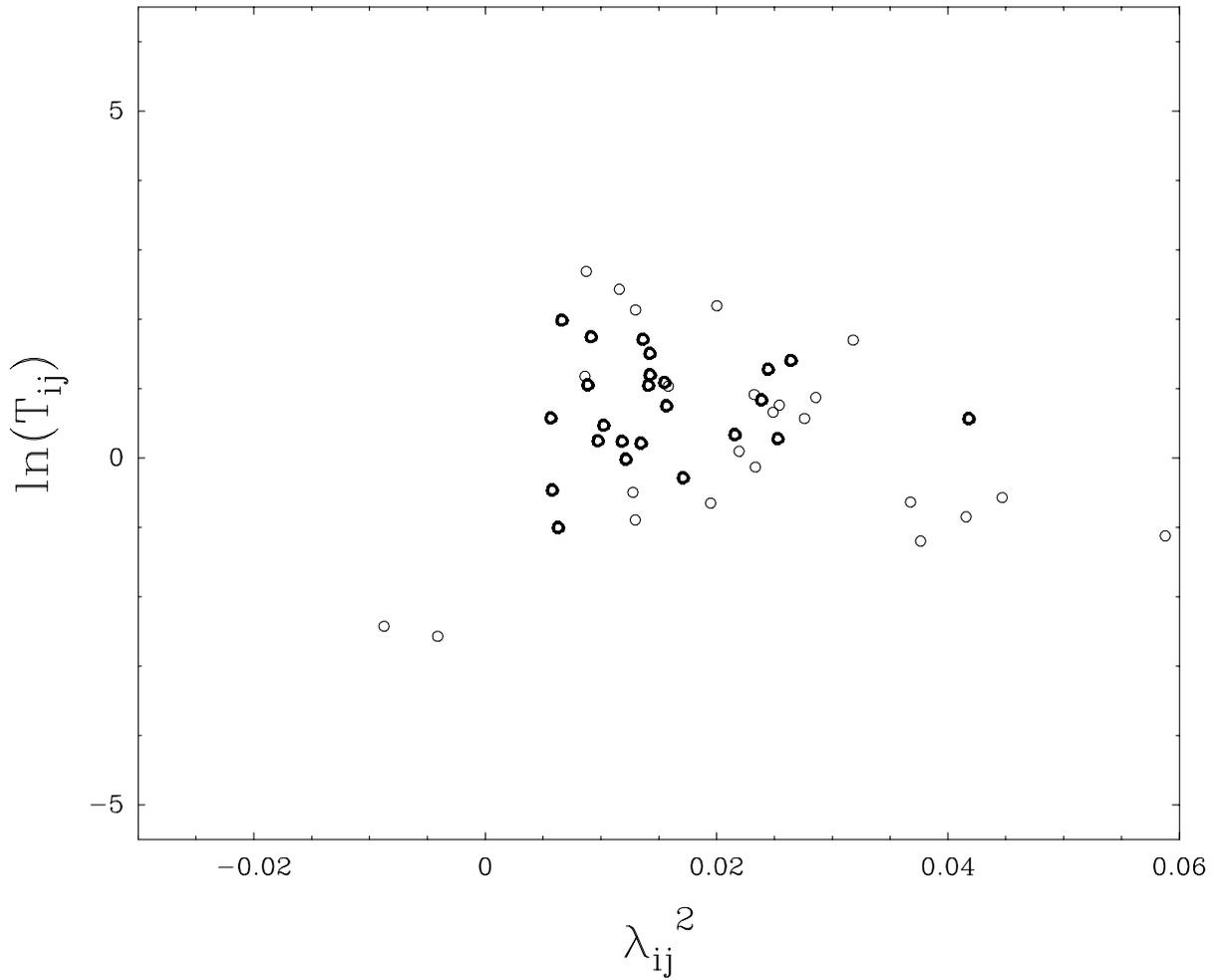, width=16cm}
\caption{\protect\label{m2} $\ln(T_{ij})$ versus $\lambda_{ij}^2$ :
each circle represents a pair of stocks taken
from the French CAC40 index (resp. from the American Dow Jones index for the
darker circles).
For both indexes, the correlation time-scales and multifractal
coefficients are approximately clustered on the same
$\lambda^2 \simeq 0.02$ and $T \simeq 1-2$ years.}
\end{center}
\end{figure}

We were not able to use intraday data
to test these predictions because the formalism requires
that the different time series should be sampled at the same times.
This can be alleviated in the future by a suitable pre-treatment that ensures
the coincidence of the sampling times.
Here, we present results for daily returns on the
magnitude covariance functions for
stocks taken from the French CAC40 index (over the
period from 1992 to 1999).
In Fig.~\ref{m1}(a), we report in dotted line the typical behavior of
$C_{ij}(\Delta t)$ versus
$\ln(\Delta t)$ for two assets of the CAC40 index
(CCF and MICHELIN). Despite the relatively poor statistical convergence,
one clearly sees a slow decay which is compatible with the logarithmic law
(\ref{vnkkd}).
From the slope and the intercept of this curve, one gets an
estimate of $\lambda^2_{ij}$ and $T_{ij}$ for these two
assets. In solid line, we have plotted the average of $C_{ij}$ performed
over all the pairs of assets in the CAC40 index.
The logarithmic
decay of the magnitude covariance is thus a remarkably stable feature among all
pairs of assets constituting the CAC40.
In Fig.~\ref{m2}(b), we show the histogram
of the values of $\lambda^2_{ij}$ estimated for 253 pairs of stocks
belonging to the
CAC40 index. Even though our estimates exhibit a large dispersion,
one clearly sees that the distribution of $\lambda^2_{ij}$ is
centered around $\lambda_{ij}^2 = 0.02$. This value, which is very
close to the estimates of $\lambda_i^2$ determined in the previous section
with the
mono-asset analysis, can
be considered as typical of financial stocks. This is confirmed
in Fig.~\ref{m2} which shows the estimates of
$\lambda^{2}_{ij}$ versus the estimates of $T_{ij}$ for pairs
of stocks taken both from the CAC40 and the Dow-Jones indices. Moreover,
one sees in Fig.~\ref{m2} that the integral time scales $T_{ij}$ are
clustered around $T=1-2$ years.

\section{Characterization of the distribution of portfolio returns}

Consider a portfolio with $n_i$ shares of asset $i$ whose initial wealth is
\be
W(0) = \sum_{i=1}^N n_i p_i(0)~.
\ee
A time $\tau$ later, the wealth has become $W(\tau) = \sum_{i=1}^N n_i
p_i(\tau)$ and the wealth variation is
\be
\delta_{\tau} W \equiv W(\tau) -W(0) = \sum_{i=1}^N n_i p_i(0)
{{p_i(\tau) - p_i(0)} \over p_i(0)}
= W(0) ~\sum_{i=1}^N w_i
\left( e^{\eta_i(\tau)}~-~1 \right) ,
\ee
where
\be
w_i = {n_i p_i(0)  \over \sum_{j=1}^{N} n_j p_j(0)}~
\ee
is the fraction in capital invested in the $i$th asset at time $0$.
Using the definition (\ref{jhghss}),
this justifies to write the return $R_{\tau}$ of the portfolio over a time
interval $\tau$ as the weighted sum of the returns $r_i(\tau)$ of the assets
$i=1,...,N$ over the time interval $\tau$
\be
R_{\tau} =
{\delta_{\tau} W \over W(0)} = \sum_{i=1}^N w_i ~r_i(\tau)~.
\label{jjkmmq}
\ee

\subsection{Cumulants of the variables $\eta_i$}

The portfolio is completely characterized
by the distribution of its returns
$R_{\tau}$ for all possible time scales $\tau$.
When the assets are assumed to
be independent, expression (\ref{jjkmmq})
shows that the pdf of $R_{\tau}$ is
obtained by the convolution of the pdf's
of the returns $r_i(\tau)$ of the
individual assets. A standard strategy
is to calculate the characteristic
functions ${\hat P_i}(k)$ of the pdf's of the returns $r_i(\tau)$,
from which one extracts the cumulants
and use their additive properties
to get the cumulants of the distribution of $R_{\tau}$. In this section 3 and
in section 4, we use the model of sections 2.1 and 2.2 of independent
multifractal
assets and turn in section 5 to the case of correlated
multifractal assets described by the formalism of section 2.3.
Let us thus estimate the cumulants of
each asset return $r_i(\tau)$ at scale $\tau$.

In that purpose, let us compute the cumulants of the ``continuous''
returns $\eta_i$'s.
The characteristic function of $\eta_i$
is obtained from Eq.(\ref{aear}) with Eq.(\ref{qkmqmmq})\,:
\be
{\hat P}_{i,\tau}(k) = \int_{-\infty}^{+\infty} du ~
{1 \over \sqrt{2\pi} ~\lambda_i(\tau)}~\exp \left( -
{\left(u-h_i(\tau)\right)^2 \over 2 \lambda_i^2(\tau)} \right)~
\exp \left( ik\mu_i T_{i} -{k^2 \over2}\sigma^2_{T_i}e^{2u} \right)~.
\label{qkmqqmmq}
\ee
From this equation, we see that the moment $m^i_n(\tau)$ of order $n$ of the
centered variable $\eta'_i(t,\tau) = \eta_i(t,\tau) - \mu_i\tau$
can be easily computed
\be
  {\partial^n {\hat P} \over \partial k^n}|_{k=0} =
   m^i_n(T_i)~
      \int_{-\infty}^{+\infty} du ~e^{nu} ~ {1 \over \sqrt{2\pi}
~\lambda_i(\tau)}~\exp \left( -
{\left(u-h_i(\tau)\right)^2 \over 2 \lambda_i^2(\tau)} \right)~, \label{jmqmq}
\ee
where $m^i_n(T_i)$ is the n-th moment of the Gaussian law at the integral
scale. The integrals in (\ref{qkmqqmmq}) and (\ref{jmqmq}) are proportional to
the Laplace transform of the normal distribution.

Using Eqs. (\ref{hi}) and (\ref{li}), we obtain the following expressions
for the moments of the centered variables $\eta_{i}'$:
\bea
    & m^i_{2n}(\tau) & = ~(2n-1)!! ~\sigma_{T_i}^{2n}
    ~\left({\tau \over T_i}\right)^{2(nh_i-\lambda_i^2n^2)}~, \label{mcp}\\
    & m^i_{2n+1}(\tau) & = ~0~. \label{mci}
\eea
Expression Eq.(\ref{mcp}) is the same as Eq.(\ref{ngwerw})
with (\ref{jnbvgaknva}) for $q=2n$.
We can then easily retrieve the moments of the (non-centered) variables
$\eta_i(t,\tau)$ by using
the definition of $\eta'$ and inverting it as
$\eta_i(t,\tau) =  \eta'_i(t,\tau) + \mu_i \tau$. Thus,
\be
\langle [\eta_i(t,\tau)]^q \rangle = \langle \left(\eta'_i(t,\tau) + \mu_i
\tau\right)^q \rangle
= \sum_{j=0}^q {q! \over (q-j)!~j!} ~ (\mu_i \tau)^{q-j}~m^i_{j}~,
\label{jfbvaknva}
\ee
for integer orders $q$.

Note that the knowledge of the moments do not allow a unique reconstruction
of the pdf. This is related to the fact that the characteristic function of the
lognormal distribution cannot be expanded in a Taylor series based on the
moments, because the moments grow too fast (as $\exp \left( n^2 \right)$) and
lead to a diverging series. Holgate [1989] has shown how to make sense of such
expansion by either using a finite Taylor series or resumming formally the
divergent moment expansion. Using the entire function or moment
constant method and Hardy's formula, one obtains [Holgate, 1989]
\be
{\hat P}_{i,\tau}(k) =
 {\rm lim}_{s \to 0}~~\sum_{j=0}^{+\infty} {(-1)^j \over 2^j} {k^{2j} \over j!}
\sigma_{T_i}^{2j}~\exp \left(2j^2 \lambda_i(\tau)^2\right)~L\left(s
e^{4j\lambda_i(\tau)^2}\right)~,
\ee
where $L$ is the Laplace transform of the lognormal distribution, which is
taken at
exponentially increasingly spaced points. Its fast decay ensures the
convergence of the series against the rapidly diverging moments and a unique
characterization of the pdf.

The first six cumulants are linked to the moments as [Ord, 1994]:
\bea
c_1 & = & m_1 ~, \label{f1a}\\
c_2 & = & m_2 - m_1^2 ~ ,\label{f1b}\\
c_3 & = & m_3 - 3m_2 m_1 + 2m_1^3 ~  ,\label{f1c}\\
c_4 & = & m_4 - 4m_3 m_1 -3 m_2^2 + 12 m_2 m_1^2 -6 m_1^4 ~ ,\label{f1d}\\
c_5 & = & m_5 - 5m_4 m_1 - 10 m_3 m_2 +20 m_3 m_1^2 + 30 m_2^2 m_1
-60m_2 m_1^3+24 m_1^5 ~  ,\label{f1e}\\
c_6 & = & m_6 - 6m_5 m_1 -15 m_4 m_2 +30 m_4 m_1^2 -10 m_3^2
+120 m_3 m_2m_1 -120 m_3 m_1^3 +30 m_2^3\nonumber\\
&& -270 m_2^2 m_1^2 +360 m_2m_1^4 -120 m_1^6 ~ .\label{f1f}
\eea
Using Eqs (\ref{mcp}), (\ref{mci}) and (\ref{jfbvaknva}),
we then determine the six first
cumulants of $\eta_i$.
Notice that the cumulants of order larger than $1$ are
invariant with respect to the translation
$\eta_i(t,\tau) =  \eta'_i(t,\tau) + \mu_i \tau$
and the centered moments (\ref{mcp}), (\ref{mci})
can thus be used directly in
expressions (\ref{f1b}-\ref{f1f}):
\bea
c_1^{i}(\tau) &=& \tau \mu_i~,    \label{cum1}\\
c_2^{i}(\tau) &=& \sigma^2_{T_i}~\left({\tau \over
T_i}\right)^{2(h_i-\lambda_i^2)}~,   \label{cum2}\\
c_3^{i}(\tau) &=& 0 ,\\
c_4^{i}(\tau) &=&  3 \sigma^4_{T_i}~
    \left({\tau \over T_i}\right)^{4(h_i-2\lambda_i^2)}\left(1-\left({\tau
\over
T_i}\right)^{4\lambda_i^2}\right)~,   \label{cum4} \\
c_5^{i}(\tau) &=& 0, \\
c_6^{i}(\tau) &=& 15 \sigma^6_{T_i}~
    \left({\tau \over
T_i}\right)^{6(h_i-3\lambda_i^2)}\left(1-3~\left({\tau \over
T_i}\right)^{8\lambda_i^2}~
 + 2~\left({\tau \over T_i}\right)^{12\lambda_i^2} \right) ~.  \label{cum6}
\eea

Expression (\ref{cum1}) retrieves
the linear dependence of the mean value as a
function
of the time interval $\tau$. This result is independent of the model.
Expression (\ref{cum2}) shows a more interesting structure: for
$h_i-\lambda_i^2 = 1/2$
we retrieve also a linear dependence of the variance
$c_2^{i}(\tau)$ as a
function of time scale $\tau$.
This linear dependence occurs in absence of or for
weak correlations of the returns. As this is a ubiquitous property
confirmed by
all empirical studies [Campbell {\it et al.}, 1997; Mantegna and Stanley,
2000], this leads to the empirical constraint
\be
h_i-\lambda_i^2 = 1/2 \label{constraint},
\ee
which is remarkably well verified (see Fig.~\ref{fig1}).

We note that, by construction of the superposition
(\ref{aear})
of Gaussian pdf's, all odd-order cumulants (except the first cumulant
corresponding to the average ``continuous'' return) are zero. The present
hierarchical
model cannot
account for possible skewness in the distribution of the ``continuous''
returns $\eta$. But as
shown in the next section, this does not implies that the skewness
of return distribution is zero. A small value of the skewness is indeed
observed in empirical studies of data sets [Campbell {\it et al.}, 1997].

If one assumes that, for all assets $h_i \simeq 0.5$, then the two
sources of statistical non-normality for $\eta$, i.e. non zero high order
cumulants, are according to our model the intermittency parameter
$\lambda_i$
and the integral time $T_i$: ``quasi-Gaussian'' statistics is recovered if
the scale $\tau$ is very close to $T_i$ or if
the intermittency parameter
$\lambda_i$ is
small enough so that the fluctuations of the variances at different scales are
small. Conversely, the distribution of ``continuous'' returns exhibits
heavy tails if
$\lambda_i$ is large or if
the time scale $\tau$ is small compared to integral scale $T_i$.

\subsection{Portfolio cumulants: perturbation expansion and
  multifractal corrections to the log-normal portfolio}

In the standard portfolio theory [Markovitz, 1959; Merton, 1990],
asset price time series are represented by
geometric Brownian motions, i.e the statistics
of ``continuous'' returns is exactly Gaussian while that of the ``discrete''
returns can be approximated by Gaussian
distributions by means of an expansion where the small
parameter is time measured in year. This results from
the definition (\ref{jhghss}) showing that the difference between
$r_i(t,\tau)$ and $\eta_i (t,\tau)$ is equal to ${1 \over 2}[\eta_i
(t,\tau)]^2$
to leading order. Since the typical magnitude of $[\eta_i (t,\tau)]^2$ is given
by the variance of $\eta_i (t,\tau)$ and since the mean and the variance
of $\eta_i (t,\tau)$ are both of the order 0.1$\tau$ (see stock examples in
Table 1) when the time $\tau$
is measured in year, $r_i(t,\tau)$ and $\eta_i (t,\tau)$ are indistinguishable
in practice for times $\tau$ of the order of or smaller than 1 year.
An alternative way of saying the same thing is that the lognormal distribution
of $r_i(t,\tau)$ reduces to its approximate Gaussian representation for
times less than or of the order of a year. Thus, the smaller
the time scale $\tau$, the more precise is the normal approximation
for $r_i(t,\tau)$ assuming a perfect Gaussian statistics for $\eta_i
(t,\tau)$ and
the smaller are high-order cumulants of $r_i(t,\tau)$, since they quantify the
departure from normality. This is the classical point of view for the
existence of non-normal returns as studied in the financial literature. This
is equivalent in our model to setting
the parameters $h_i = 0.5$ and $\lambda_i^2 = 0$.

Non-trivial multifractal corrections arise when $\lambda_i^2$ is not
negligible and thus $h_i$ departs from $0.5$. For the statistical properties of
(both ``discrete'' and ``continuous'') returns,
these multifractal corrections to the log-normal picture are stronger at fine
scales: the finer the scale, the larger the values
of high order cumulants. With respect to time scales,
this phenomenon thus acts in the way opposite
to the previous effect of the difference between
``discrete'' returns $r_i(t,\tau)$ and ``continuous'' returns $\eta_i
(t,\tau)$.
Due to multifractality, we do not expect to observe normality for returns
at small time scales, in contradiction with the standard geometrical
Brownian model.

To sum up, due to the difference between
``discrete'' returns $r_i(t,\tau)$ and ``continuous'' returns $\eta_i
(t,\tau)$,
normality is expected at small scales while deviations from normality is
predicted at large time scales. Due to multifractality, non-normality is
expected at small scale and Gaussian statistics are predicted at large time
scales,
beyond the ``integral'' scale. Putting the two effects together, non-normality
must be the rule at all time scales, with a cross-over from small to large
time scales,
from non-normality induced by multifractility to
non-normality induced by the difference between continuous and
discrete returns. Thus, we expect high order return cumulants to be non
zero both
at large scales because of the log-normality and also at small
scales because of multifractality of $\eta$.

Let us now quantify these effects. In this purpose,
we compute the cumulants of the portfolio returns using an
expansion where the small parameter is $\tau$ (measured in year),
which takes multifractal corrections into account.
From Eq.(\ref{jjkmmq}), the cumulants $C_n^P(\tau)$ of the portfolio
can be expressed in terms of the cumulants $C^i_n(\tau)$ of the
returns for each individual asset as weighted sums over them:
\bea
  & C_1^P(\tau) & = \sum_{i=1}^N ~w_i ~C^i_1 (\tau) ~, \\
  & C_n^P(\tau) & = \sum_{i=1}^N ~w_i^n ~C_n^i(\tau) ~,~~~~~~~~ \mbox{for}
~ n > 1~.
\eea

Using Eqs. (\ref{cum1}-\ref{constraint}) and from the definition
$r(\tau) = e^\eta-1$, after some algebra we find, to
the leading orders in the time expansion,
\bea
C_1^{P}(\tau) &\simeq & \sum_{i=1}^N w_i \tau (\mu_i+{1 \over
2}\sigma_i^2)~,    \label{rcum1}\\
C_2^{P}(\tau) &\simeq & \sum_{i=1}^N w_i^2 \sigma^2_{i} \tau~,
\label{rcum2}\\
C_3^{P}(\tau) &\simeq& \sum_{i=1}^N w_i^3 \tau^2
\left(3\sigma_i^4-18\sigma_{i}^4 \lambda_i^2 \ln({\tau \over T_i})
\right), \label{rcum3}\\
C_4^{P}(\tau) &\simeq&  \sum_{i=1}^N w_i^4
\left(16\sigma_i^6\tau^3-12\sigma_i^4\lambda_i^2 \tau^{2}
\ln({\tau \over T_i}) \right),   \label{rcum4} \\
C_5^{P}(\tau) &\simeq& \sum_{i=1}^N w_i^5
\left(125\sigma_i^8\tau^4-240\sigma_i^6 \lambda_i^2
\tau^3\ln({\tau \over T_i}) \right), \label{rcum5}\\
C_6^{P}(\tau) &\simeq& \sum_{i=1}^N w_i^6
\left(1296\sigma_i^{10}\tau^5+720\sigma_i^6\lambda_i^4 \tau^3
\ln^{2}({\tau \over T_i}) \right) \; , \label{rcum6}
\eea
where we have set $\sigma_i^2=\sigma_{T_i}^2/T_i$ (in year$^{-1}$).
We can see how the multifractal corrections arise in the high order
cumulants. Using the typical values $\mu_i \sim \sigma_i \sim \lambda_i=0.1$
and $T_i=1$, one can compute, according to the value of $\tau$,
which term from the log-normal expansion or from the multifractal
correction dominates the cumulant. For the first two cumulants $C_1$ and
$C_2$, it
can be shown that multifracal corrections are always negligible.
For the third-order cumulant $C_3$, for time scales larger than a few seconds,
the multifractal
nature of $\eta$ is not important. However, for cumulants of order
larger than or equal to 4, the multifractal corrections are very important
for time scales as large as several months: for instance,
using the previous values for the parameters, we find that the
multifractal term is dominating for all times $\tau$ smaller
than $\tau^{\ast} \approx$ 6 months.

From the structure of the cumulants given by Eqs
(\ref{cum2},\ref{cum4},\ref{cum6}) and
their generalization to higher order, it is easy to show by recurrence that
the multifractal correction to the
$2n$-th order cumulant behaves, for $n > 1$, as
\be
C^{P}_{2n}(\tau) \simeq \sum_{i=1}^N w_i^{2n} \sigma_{i}^{2n}~\tau^n ~
(-1)^{n-1} ~\lambda^{2(n-1)}_i ~\ln^{n-1} \left({\tau \over T_i}\right)~,
\label{dgzbjz}
\ee
to leading order.
As a consequence, the corrections
to the Gaussian description brought by the
multifractal model are all the more important than
the order of the cumulant is higher, i.e. than one looks
further in the tail of the return distribution.
We retrieve a common observation
that the deviations from the Gaussian model
are all the more important for
large risks
quantified by the behavior of the tails
of the portfolio return distribution
[Sornette, 1998; Sornette {\it et al.}, 2000].

If we neglect the higher order cumulants in the previous expansion, we
retrieve that
the portfolio return $C_1^{P}(\tau) $ and variance
$C_2^{P}(\tau)$ are both
proportional to time scale $\tau$.
Hence, the portfolio optimization of the asset
weights $w_i$ are independent of the investment horizon $\tau$ in the
absence of the riskless
asset. This is the
well-known
result underlying Gaussian portfolio optimization
[Markovitz, 1959; Merton, 1990].
In contrast, in the hierarchical model,
cumulants of order 4 and higher have a different dependence
on the investment time horizon $\tau$. This implies that large risks exhibit
a non-trivial time dependence: it will therefore not be possible to
optimize the asset weights $w_i$
for all time scales $\tau$ simultaneously. This is the novel ingredient
captured by our hierarchical model: a portfolio optimization
which is concerned with large risks has to optimize its
investment time horizon in a manner that we are now
going to investigate.

\subsection{Excess kurtosis}

\begin{figure}
\epsfig{file=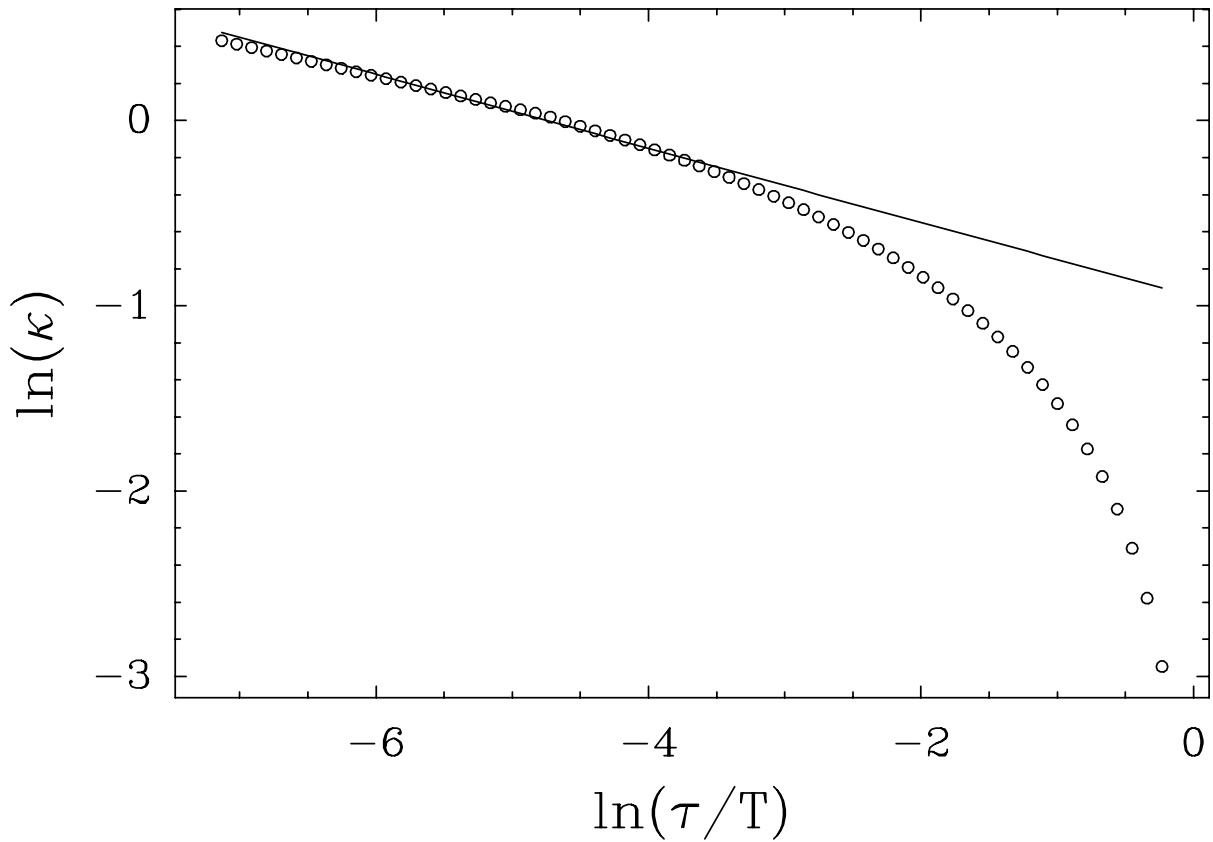, width=16cm}
\caption{\protect\label{kurtopredc}
Excess kurtosis $\kappa$ given by the prediction (\ref{jghlala}) of the
multifractal cascade model with parameters $\lambda^{2}=0.018$ and
$\sigma^{2}=1.8~10^{-4}$ (see S\&P500 data in Table 1) in logarithmic scale
as a function of $\tau/T$ also in logarithmic scale. The decay is not a pure
straight line but can be approximated by one with an average exponent close to
$0.2$ for small $\tau$'s as previously observed [Dacorogna {\it et al.}, 1993;
Ding {\it et al.}, 1993; Bouchaud {\it et al.}, 2000].}
\end{figure}

The absence of volatility correlations and the validity of the
log-normal model would imply, as already said, that
the variance $\lambda_i^2$ of the volatilities vanishes.
As seen from expressions
(\ref{rcum4}-\ref{rcum6}), all cumulants of order $n$
larger than $2$ would then be of order $\sigma_i^{2(n-1)}\tau^{n-1}$, i.e.
very small at small time scales, as
expected for a Gaussian distribution of returns.
Actually, for instance for the  US S\&P500 index,
there is a residual hierarchical correlation structure,
quantified by
$h=0.518$ and $\lambda^2 =0.018$ and $T \simeq 1$ year.
With these values,
we predict an excess kurtosis by using Eq. (\ref{rcum2})
and (\ref{rcum4}) for a single
asset (the S\&P500 index)
\be
\kappa \equiv {C_4 \over C_2^2} = 16\sigma^2\tau-12\lambda^2\ln(\tau) \; .
\label{jghlala}
\ee
This expression makes clear that the size of the variance $\lambda^2 \neq
0$ of the volatilities at
different scales quantifies the distance from the log-normal
paradigm. The excess kurtosis at small time scale is negligible
for $\lambda^2 = 0$.
In contrast, in the presence of multifractality ($\lambda^2 \neq 0$),
the excess kurtosis $\kappa$ is predicted to be very large
at small time scales. The dependence of the excess kurtosis is shown in Fig.
\ref{kurtopredc}. This figure shows
that the excess kurtosis given by Eq. (\ref{jghlala}) can be represented
approximately as
a power law decay $\kappa \sim (\tau/T)^{-0.2}$ over more than three decade
with a small exponent $\approx 0.2$, in excellent
agreement with previous determinations [Dacorogna {\it et al.}, 1993;
Ding {\it et al.}, 1993; Bouchaud {\it et al.}, 2000].
To confirm the relevance of Eq. (\ref{jghlala}), one thus needs to investigate
the excess kurtosis for time scales large enough as compared to the integral
scale $T$. It is important to contrast this
decay of
the excess kurtosis $\kappa$ as a function of the time scale $\tau$ with what
one would expect from a model without correlations across scales: in that case,
all cumulants are linear in $\tau$ and $\kappa \propto 1/\tau$. The
anomalous law
(\ref{jghlala}) is a clear signature of long-range correlations in the
volatilities.

\section{Portfolio optimization with time-scale dependent risks}

\subsection{The expected utility approach}

Starting the period with initial capital $W_0>0$, the investor is assumed
to have preferences that are rational in the von-Neumann-Morgenstern
[1944] sense
with respect to the end-of-period distribution of wealth $W_0 + \delta S$.
His preferences are
therefore representable by a utility function $u(W_0 + \delta S)$
determined by the wealth
variation $\delta S$ at the end-of-period $\tau$. The expected utility theorem
states that the investor's problem is to maximize
$\E[u(W_0 + \delta S)]$, where ${\rm E}[x]$ denotes the expectation
operator\,:
\be
{\rm E}^{\tau}[u(W_0 + \delta S)] = \int_{-W_0}^{+\infty} d\delta S~
u(W_0 + \delta S) ~P^{\tau}_S(\delta S)~.
\label{hqlql}
\ee
The utility function $u(W)$ has a positive first derivative (wealth is
prefered) and a negative
second derivative (risk aversion). Use of the utility maximization in
portfolio optimization
can be found in [Levy and Markowitz, 1979; Kroll {\it et al.}, 1984].

Here, we consider a simple case of a constant absolute measure of
risk aversion $-u''/u' = a$ (where the primes
denote the derivatives), for which $u(W) = -\exp(-a W)$.
With a very good approximation for large initial wealths,
we can take $P^{\tau}_{S}(\delta S<-W_0) \simeq 0$. This gives
\be
{\rm E}^{\tau}[u(W_0+\delta S)] = - e^{-aW_0} \int_{-\infty}^{+\infty}
e^{-a \delta S} P^{\tau}_{S}(\delta S) d(\delta S)~,
\ee
which is nothing but the Fourier transform of the probability distribution
function
(with imaginary argument $k$):
\be
{\rm E}^{\tau}[u(W_0+\delta S)] = - e^{-aW_0} \hat{P^{\tau}_{S}}(k = ia)~.
\ee
By definition of the cumulants, this reads
\be
{\rm E}^{\tau}[u(W_0+\delta S)] = - e^{-aW_0} ~\exp\left(-a C_1^{P}(\tau) +
\sum_{n=2}^{+\infty}
{(-a)^{n} \over (n)!} C_{n}^{P}(\tau)\right)~.   \label{jfjakka}
\ee
Maximizing the expected utility
${\rm E}^{\tau}[u(W_0+\delta S)]$ thus amounts to minimizing the argument
$-a C_1^{P}(\tau) + \sum_{n=2}^{+\infty} {(-a)^{n} \over
(n)!} C_{n}^{P}(\tau)$ of the second exponential in the r.h.s. of
(\ref{jfjakka})
with respect to the weights $w_i$.

Keeping in mind that the time $\tau$ is the small parameter of the
problem, we express the optimal asset weights
$w_i$ as equal to the optimal Markowitz results
denoted $w^o_i$ (obtained by droping all multifractal corrections)
plus the multifractal corrections. 
Keeping only the cumulants of order up to 4
(this approximation is valid not only
in the small time limit but for a risk aversion $a$ not too large).
The problem is thus to minimize
\bea
&-& \alpha \sum_{i=1}^N w_i - a \sum_{i=1}^N w_i \tau (\mu_i+{1 \over
2}\sigma_i^2)~
+ {a^2 \over 2} \sum_{i=1}^N w_i^2~\tau~\sigma^2_{i}
\nonumber \\
&-& {a^3 \over 6} \sum_{i=1}^N w_i^3 \tau^2 3\sigma_i^4 
+ {a^4 \over 24} \sum_{i=1}^N w_i^4 
\left(16\sigma_i^6\tau^{3}-12\sigma_i^4\lambda_i^2\tau^2\ln({\tau \over T_i})\right)
\eea
with respect to the asset weights $w_i$'s. The term $\alpha$ is a Lagrange
multiplier
ensuring the normalization $\sum_{i=1}^N w_i = 1$.
This optimization amounts to finding the roots of a third order
polynomial and one can thus obtain closed expressions for the
weights $w_i$. 

In order to quantify the influence of multifractal
corrections and to handle simple expressions, let us neglect
the corrections due to the difference between ``continuous'' and ``discrete''
returns which are unimportant at small time scales up to 6 months.
In this approximation, the problem is to minimize
\be
- \alpha \sum_{i=1}^N w_i - a \sum_{i=1}^N w_i \tau \mu_i+~
+ {a^2 \over 2} \sum_{i=1}^N w_i^2~\tau~\sigma^2_{i}
- {a^4 \over 24} \sum_{i=1}^N w_i^4 12\sigma_i^4\lambda_i^2\tau^2\ln({\tau
\over T_i}).
\ee
Then, in absence of the last terms proportional to $a^4$, i.e. for all
$\lambda_i^2$'s equal to zero, we get the ``Markowitz'' solution
\be
w^o_j = {a \mu_j\tau + \alpha \over a^2
\sigma^2_{j}~\tau}~,
\label{fgkanak}
\ee
where $\alpha$ is determined from the normalization condition.
In the simple case where $\sigma^2_{i} = \sigma^2$ is the same
for all assets, this gives
\be
w^o_j = {1 \over N} + {1 \over a \sigma^2}(\mu_j - \langle \mu \rangle)~,
\label{bvbzkzm}
\ee
where $\langle \mu \rangle$ is the mean return averaged over all assets:
assets with better
than average returns have thus more weight in the portfolio, with
a leverage controlled by the risk $\sigma^2$.

Using this solution (\ref{fgkanak}), we get the general solution of the
weights $w_i$ up to second order in powers of $\lambda_i^2$
\be
w_i  = w^o_i ~\left[1 + A_i \lambda_i^2 \right] ~, \label{jfksaka}
\ee
where
\be
A_{i} = 2 a^{2} \sigma_{i}^{2} w_{i}^{0} \tau \ln ({\tau \over T_{i}})~.
\ee
The expression of the weigths valid to first-order in the multifractal
corrections for $\tau \leq \tau^{\ast}$ as a function of time horizon
$\tau$ is then
\be
w_i = w_{i}^{0}~\left[1+2 a^{2} \sigma_{i}^{2} w_{i}^{0} \lambda_{i}^{2}
\tau \ln ({\tau \over T_{i}})  \right] ~.
\ee

Returning to the general solution (\ref{jfksaka}), we see that assets with
integral time scales $T_i \approx e \tau$ (where $e \approx 2.718$ is the
base of the natural logarithm)
will be the most depleted compared to the Gaussian
solution, as the factor $A_i$ is negative with a maximum amplitude for
$\tau/T_{i}=1/e$. Both
for small $\tau/T_i$ and for
$\tau/T_i$ approaching $\tau^{\ast}$, the solution is close to the Gaussian
solution as it should : small $\tau/T_i$
do not lead to large absolute risks; $\tau/T_i \approx \tau^{\ast}$ leads to
the Gaussian regime $\tau/T_i = 1$ for $\eta_{i}$.
The worst case occurs for intermediate values of the time horizon compared
to the integral time scale.
Such stocks will be unfavored in the portfolio selection.

This result is actually more general
than this section would lead us to believe. This can be seen from the
structure of the cumulants of order $2n$
given by Eq.
(\ref{dgzbjz}): the dependence in $\tau/T_i$
given by $\left({\tau \over T_i}\right)^n \left(-\ln \left({\tau \over
T_i}\right)\right)^{n-1}$
shows that, for large $n$, the cumulants are the largest when
$|(\tau/T_i)\ln (\tau/T_i)|$ is the largest.
This is exactly the same term that controls the corrections to the Gaussian
case quantified by the
parameters $A_i$.

\subsection{Efficient frontiers for multi-period portfolio optimization}

As recalled in the introduction, the multiperiod portfolio problem
as been addressed by several studies. This problem is a
natural application of our multi-scale description of returns.
The problem we investigate is to minimize a risk measure represented by
a cumulant of order $2n$ for a fixed mean return.
For instance,
one can choose to minimize the value of the cumulant $C_{4}$
at fixed mean return $C_1$ (this amounts to find the set of $w_{i}$'s
that minimize the pseudo-utility function
$-a C^{P}_{1}+C^{P}_{4} = -a \sum_{i=1}^N w_{i} C^{i}_{1}+\sum_{i=1}^N
w^{4}_{i} C^{i}_{4}$), thus defining the $C_1-C_4$
efficient frontier [Andersen and Sornette, 2000].
The previous cumulant perturbative expansion
at different time scales $\tau$ allows us to
estimate the shape of the generalized optimal frontiers for
all periods, given a fixed horizon $T_h$.

For the sake of simplicity,
we will suppose that all the assets are characterized
by the same
integral time values $T_i=T=1$~year,
which is a reasonable value as revealed by
the empirical analysis of section 2.
Denoting $T_h$ the investment horizon and $N_p$ the number of periods, we will
consider the portfolio returns at scale $\tau=T_h/N_p$.
Moreover,
we will explicitely make the so-called ``rebalancing assumption'':
periodic rebalancing supposes that, at
the end of each period, portfolio composition is adjusted in
order to restore the original weights.
We will assume that doing so the returns associated to each period
are statistically independent.

\begin{figure}
\label{figrt}
\begin{center}
\epsfig{file=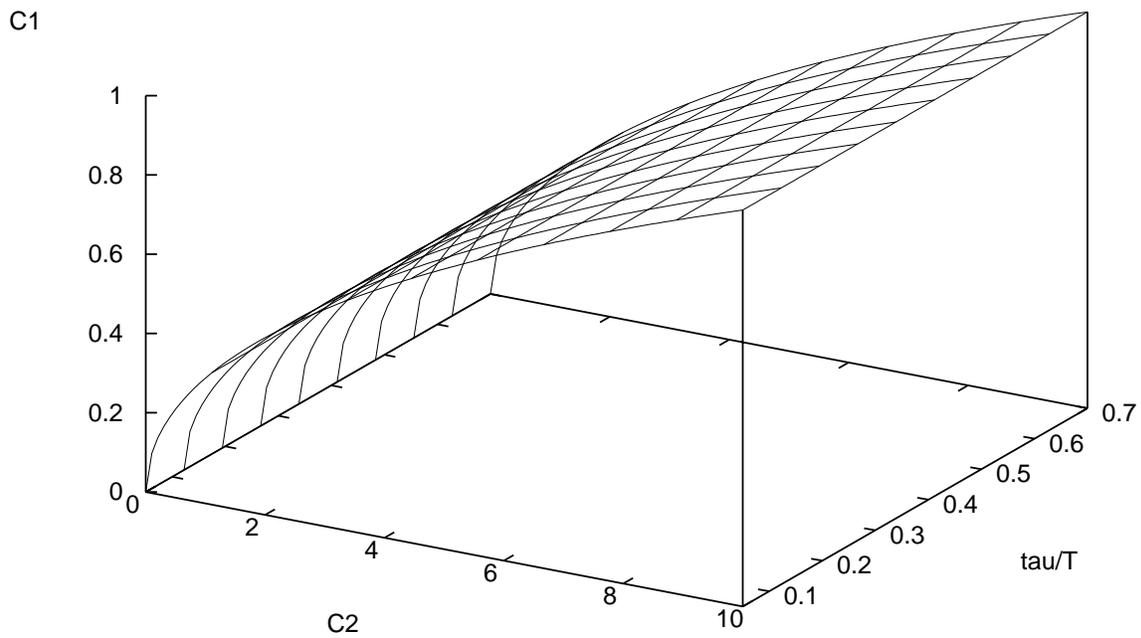, width=12cm, angle=-90}
\caption{\protect\label{cp2cp1} Standard efficient frontier
  for multiperiod portfolio:
$C_1^{P}(T_h,\tau)$ as a function of $C_2^{P}(T_h,\tau)$ for various
time-scales $\tau$'s. Units are arbitrary. The shape of
the frontier illustrates the independence with respect to the time horizon.
}
\end{center}
\end{figure}

Once again, in order to quantify the influence of multifractal
corrections and to handle simple expressions, we will neglect here
the corrections due to the difference between ``continuous'' and ``discrete''
returns which are unimportant at small time scales up to 6 months (above
which they are dominated by the log-normal corrections).
One can easily show using $r_{i} \simeq \eta_{i}$ and the ``rebalancing
assumption'' that the cumulants of the portfolio returns
for a given multiperiod strategy have exactly the same
expression as in Eqs. (\ref{rcum1})-(\ref{rcum6}) where
$\mu_i \tau$ and $\sigma_i^{2n} \tau^{n}$ become respectively $\mu_i T_h$ and
$\sigma_i^{2n} T_h \tau^{(n-1)}$ and the multifractal corrections
$\lambda_i^{2(n-1)}
\tau^{n} \ln^{(n-1)}(\tau)$ become $T_{h} \lambda_i^{2(n-1)}
\tau^{(n-1)} \ln^{(n-1)}(\tau)$.
Using this rule, it is easy to show that, if one estimates the
risk using the variance of portfolio returns, the results
are independent of the number of periods at fixed horizon $T_h$.
Fig.~\ref{cp2cp1} represents the standard efficient frontier
$C_1^{P}(T_h,\tau)$ as a function of $C_2^{P}(T_h,\tau)$ for various
$\tau$'s. As expected, it is clearly
apparent that the efficient frontiers are {\it independent} of time
$\tau$, i.e. of the number of periods.
This is nothing but Tobin's result [1965] that the
single-period minimum
variance set and the multi-period minimum variance set are identical.

\begin{figure}
\label{figer}
\begin{center}
\epsfig{file=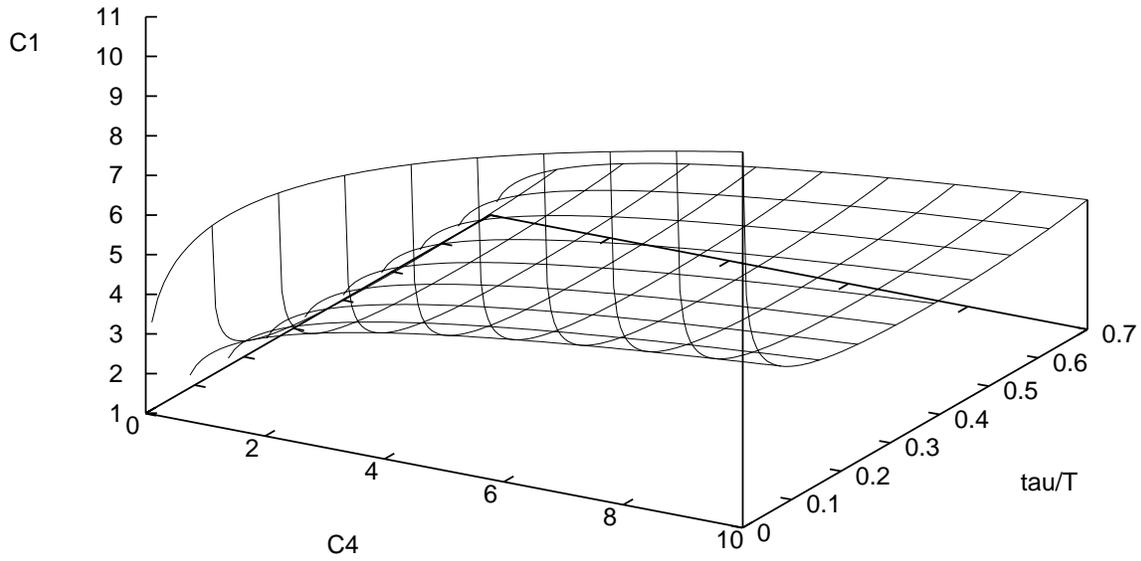, width=12cm, angle=-90}
\caption{\protect\label{cp4cp1} Generalized multiperiod efficient frontiers
$C_1^{P}(T_h,\tau)$ as a function of $C_4^{P}(T_h,\tau)$ for various
$\tau/T$'s in the case of a single risky asset. The return of the unrisky
asset has be chosen to be $\mu_0=0.05$ and the parameters of the
risky asset are $\mu=0.15$, $\sigma=0.1$, $T=1$ year and $\lambda^2=0.02$.
$T_h$ is set to 1 year.
}
\end{center}
\end{figure}

In contrast, the generalized
efficient frontiers with a risk measure represented by
higher order cumulants will involve
multifractal corrections and both portfolio composition and efficient
frontiers will depend on the number of periods.
Let us illustrate this results for the $C_4^{P}$ minimization
problem. In that case, the optimum weight of asset $i$
can be written for $\tau \leq \tau^{\ast}$ as:
\be
   w_i = \left( { a \mu_i \over
       - 48 \sigma_i^4 \lambda_i^2 \tau \ln({\tau \over T_{i}})}
     \right)^{{1 \over 3}} \; ,
\ee
where $a$ is the Lagrange parameter associated with the fixed
mean return.
For a large number of periods (i.e. small time
scale $\tau$), assets with small multifractal
parameter $\lambda_i^2$ are prefered.
Using these weights,
Fig. \ref{cp4cp1} shows the efficient frontiers
$C_1^{P}(T_h,\tau)$ as a
function of $C_4^{P}(T_h,\tau)$
for various periods.
We have considered the simplest case of a single risky
asset with parameters $\mu$ for the mean return, $\sigma$ for the
standard deviation and $\lambda^2$ for the multifractal parameter.
Denoting $\mu_0$ the return
of the riskless asset, the parametric equation for
the efficient frontier for an horizon $T_h$ and a
number of periods $T_h/\tau$ is:
\bea
  C_1^P(T_h,\tau) &=& \mu_0 T_h +\left(a (\mu-\mu_0)
      \over -48\sigma^4\lambda^2\tau\ln({\tau \over T}) \right)^{{1
\over 3}}(\mu-\mu_0)T_h ~, \\
  C_4^P(T_h,\tau) &=& - \left({a (\mu-\mu_0)
      \over -48\sigma^4\lambda^2\tau\ln({\tau \over T})} \right)^{{4
\over 3}} 12T_h\sigma^2\lambda^2\tau\ln({\tau \over T}).
\eea
Fig. \ref{cp4cp1} exhibits a
measurable dependence on the time scale $\tau$, in contrast with the standard
mean-variance efficient frontier $C_1^P(T_h,\tau)$ as a function of
$C_2^P(T_h,\tau)$.
For a fixed risk $C_4^P$, the return is seen to be increasing at both ends
as a function of time scale $\tau/T$, i.e., for $\tau/T \to 0$ and for
$\tau/T \to \tau{\ast}$.
There is thus a worst choice for the time-horizon and for the rebalancing
of the portfolio, approximately given by one fifth of the integral time scale
$T$. For this time scale, the return is minimum for a given (large) risk and
the risk is maximum for a given return. One the other hand one can see that
the shorter the rebalancing period, the lower is the risk for a given return.
As a matter of facts, the risk is here quantified by the fourth cumulant which
goes to zero as $\tau \to 0$ (Gaussian statistics). The limiting factor for
small $\tau$'s is then transaction cost. Fig. 8 makes more precise this effect
by plotting $C_1^P(T_h,\tau)$ as a function of $\tau/T$ for a given value $C_4^P=2$ of
the accepted risk.

\section{Distribution of returns of a portfolio of correlated
assets within the multifractal multivariate cascade framework}

In order to discuss portfolio optimization in presence
of correlated assets, let us make the approximation
(valid for $\tau$ small enough) that $\eta_i(\tau) \simeq r_i(\tau)$.
Moreover, for the sake of simplicity,
let us assume that the $\eta_i$'s are centered, i.e., $\mu_i = 0$.
Using the notations of section 2.3,
the characteristic function ${\hat P}(k)$
of the portfolio $R(\tau)$ can be written as
\be
  {\hat P}_P(k)~=~\int_{-\infty}^{+\infty} d^Nu \;
\cN_N\left(\vu-\vh(\tau),\mL(\tau)\right) \;
   \int_{-\infty}^{+\infty} d^Nr \; \cN_N\left(\vr,\mD(\vu) \ms_0
\mD(\vu)\right) e^{ik{\vec w}.\vr} \; ,
\ee
where we have denoted $\vec w$ the vector of the weights $w_i$.
Let us consider the orthogonal matrices $\mQ$ and $\mO$ that
diagonalize respectively the matrices $\mL$ (and thus also the matrix
$\lambda^2_{ij}$ by virtue of Eq. (\ref{ns}), showing that $\mQ$ is
independent of $\tau$)
and $\ms_0$:
\bea
 \mL(\tau) & = & \mQt {\tilde \mL}(\tau) \mQ ~,\\
 \ms_0       & = & \mOt_0 {\tilde \ms_0} \mO_0~,
\eea
where $\tilde \mL$ and $\tilde \ms_0$ are the
diagonal matrices formed of the eigenvalues of $\mL$ and $\ms_0$.
Let us set ${\vec r}' = \mO \mD^{-1}(\vu)\vr$. If we denote ${\vec w}'$ the
vector $\mO{\vec w}$,
the previous equation can be rewritten as:
\be
  {\hat P}_P(k) ~=~\int_{-\infty}^{+\infty} d^Nu \;
\cN_N\left(\vu-\vh(\tau),\mL(\tau)\right) \;
  \int_{-\infty}^{+\infty} d^Nr' \prod_{i=1}^N \cN_1(r'_i,\sigma^2_{T_i})
e^{ike^{u_i}w_i'r_i'} \; ,
\ee
where $\sigma^2_{T_i}$ are the eigenvalues of $\tilde \ms_0$ and $\cN_1$ is the
standard monovariate Gaussian law. After calculating the second integral,
we obtain
the following expression for the
characteristic function:
\be
  {\hat P}_P(k) ~=~\int_{-\infty}^{+\infty} d^Nu \;
\cN_N\left(\vu-\vh(\tau),\mL(\tau)\right) \;
    e^{-{k^2 \over 2}\sum_{i=1}^N e^{2u_i} {w'}^2_i \sigma^2_{T_i}}\; .
\ee

From this equation, one can immediately see that, if all the cascades are
uncorrelated, i.e, $\mL$
is diagonal, the characteristic function is the same as in the case of
independent assets
where the weights $w_i$ have been replaced by the weights $w'_i$. This is
the same situation
as in classical (Gaussian) portfolio theory where the case of correlated
assets is reduced
to the uncorrelated one by such a simple change of variable on the weights.

When the covariance matrix $\mL$ is non trivial, the situation is more
complicated.
Let us consider the case where each matrix $\mL_l$ in the block
decomposition (\ref{bd})
is singular, i.e, $[\mL_l]_{ij} = \lambda^2_l$. This case corresponds to
the existence of $L$ degrees of freedom in the problem for which one can
compute the characteristic function under the form:
\bea
  {\hat P}_P(k) &=& \int_{-\infty}^{+\infty} du_1...du_L~
  \prod_{l=1}^L  \cN_1\left({u_l-h_l \ln(\tau/T_l) \over \lambda_l^2
\ln(\tau/T_l)}\right) ~
    e^{-{k^2 \over 2} N_l w_l'\sigma^2_{T_l}e^{2u_l}}~, \nonumber  \\
    &=& \prod_{l=1}^L \left( \int_{-\infty}^{+\infty} du_l
    \cN_1\left({u_l-h_l \ln(\tau/T_l) \over \lambda_l^2
\ln(\tau/T_l)}\right) ~
    e^{-{k^2 \over 2} N_l w_l'\sigma^2_{T_l}e^{2u_l}} \right)~.
\label{nvbbzx}
\eea
This case corresponds to $L$ independent assets whose integral
scale variances are $N_l \sigma^2_{T_l}$.

Expression (\ref{nvbbzx}) is the product of $L$ terms of the form
(\ref{qkmqqmmq})
already encountered for the case of a single asset in section 3. The
moments and
cumulants can thus be calculated analytically.

\section{Conclusion}

We have extended a statistical model of price returns and
volatility correlations
based on the idea of information cascades from large time scales to smaller
time scales. Empirical tests performed on intra-day as well as daily data
on the CAC40 and
S\&P500 indices, on their constituting stocks as well as on bonds and
currencies
 validate satisfactorily the
model. The calibration give a robust and
seemingly consistent
value for the two key parameters:
 the integral time-scale is found in the range of one to two years and the
variance of
the multiplicative kernel is approximately $0.02$ for all stocks and
indices that have been investigated. Our results show that the evidence for
multifractality is fully consistent with a simple cascade origin, flowing
from large time-scales
to shorter time-scales. The multifractal cascade model offers an intuitive
explanation
for the observation that the detection of ``abnormal'' states or crises in
the stock market
requires an index constructed over many different horizon times [Zumbach et
al., 2000].

We have also offered an extension of the cascade model into a multi-variate
framework to
account for correlations between assets, in addition to the correlations in
time-scales.
Future works will exploit this novel formalism, in particular for multi-period
portfolio characterization and optimization.

In a second part, we have shown how to characterize the distribution of returns
of a portfolio constituted of assets with returns distributed according to
such
multifractal cascade distributions. In particular, explicit analytical
expressions
for the first six cumulants are offered. We also show that, within a
utility approach
with a constant absolute measure of risk aversion (exponential utility
function),
the problem of portfolio optimization
amounts to maximize a sum over cumulants weighted by powers of the risk
aversion coefficient.
Working in the space of (return, fourth-order cumulant) or of (return,
sixth-order cumulant)
generalizes the mean-variance approach and underlines the impact of the
investor horizon-time. The most important consequence of the theory is that
the optimal portfolio depends on the time-scale. In addition, it is not
possible
to simultaneously optimise all the components of risks with respect to the
choice
of the investment time scale. This result extends to the investment horizon
dimension
previous results obtained from a decomposition of the risk into a spectrum
from small to
large risks quantified by the cumulants of the distribution of portfolio
returns
[Sornette {\it et al.}, 2000].

In principle, our portfolio theory allows one to quantify how much
diversification can be obtained by
buying different assets and managing them optimally with respect to their
possibly different integral time scales: indeed, our results
suggest that reallocation of assets in the portfolio
should be performed with different time-horizon depending upon the assets.
This is related to the ``time-diversification'' concept introduced by
Martellini [2000]
for option hedging in the presence of transaction costs. A quantification
and tests
of these strategies will be reported elsewhere.

\pagebreak

REFERENCES\,:

J.V. Andersen and D. Sornette,
Have your cake and eat it too: increasing returns while lowering large risks!
preprint at http://xxx.lanl.gov/abs/cond-mat/9907217.

F.D. Arditti and H. Levy,
Portfolio efficiency analysis in three moments - the multiperiod case,
Journal of Finance 30, 797-809 (1975).

A. Arneodo, J.F. Muzy and D. Sornette,
``Direct'' causal cascade in the stock market, Eur. Phys. J. B
2, 277-282 (1998a).

A. Arneodo, E. Bacry, S. Manneville and J.F. Muzy,
Analysis of random cascades using space-scale correlation functions,
Phys. Rev. Lett. 80, 708-711 (1998b).

A. Arneodo, S. Manneville and J.F. Muzy,
Towards log-normal statistics in high Reynolds number turbulence,
Eur. Phys. J. B 1, 129-140 (1998c).

A. Arneodo, S. Manneville, J.F. Muzy and S.G. Roux,
Revealing a lognormal cascading process in turbulent velocity statistics
with wavelet analysis, Phil. Trans. Royal Soc. London Series A 357,
2415-2438 (1999).

G. Ballocchi, M. M. Dacorogna and R. Gencay,
Intraday statistical properties of Eurofutures by Barbara Piccinato,
Derivatives Quarterly 6 (2), 28-44 (1999).

A. Bershadskii, Multifractal critical phenomena in traffic and economic
processes, Eur. Phys. J. B 11, 361-364 (1999).

H. Bierman, Jr., Portfolio allocation and the investment horizon,
Journal of Portfolio Management 23, 51-55 (1997).

H. Bierman, Jr., A utility approach to the portfolio allocation decision
and the investment horizon, Journal of Portfolio Management 25, 81-87 (1998).

J.-P. Bouchaud, M. Potters and M. Meyer,
Apparent multifractality in financial time series,
Eur. Phys. J. B 13, 595-599 (2000).

M.-E. Brachet, E. Taflin and J. M. Tcheou,
Scaling transformation and probability distributions for financial time series,
preprint cond-mat/9905169 (1999).

W. Breymann, S. Ghashghaie and P. Talkner, A stochastic cascade model for
FX dynamics, preprint cond-mat/0004179 (2000).

J.Y. Campbell, A.W. Lo and A.C. MacKinlay,
The econometrics of financial markets (Princeton University Press,
Princeton, New Jersey, 1997).

B. Castaing, Y. Gagne and E.J. Hopfinger, Velocity probability density
functions of high Reynolds number turbulence, Physica D 46, 177-200 (1990).

B. Chabaud, A. Naert, J. Peinke, F. Chilla and B. Castaing, Transition towards
developed turbulence, Phys. Rev. Lett. 73, 3227-3230 (1994).

M.M. Dacorogna, U.A. M\"{u}ller, R.J. Nagler, R.B. Olsen {\it et al.},
 A geographical model for the daily and weekly seasonal volatility in the
 foreign exchange market, Journal of International Money \& Finance 12,
413-438 (1993).

M.M. Dacorogna, U.A. M\"{u}ller, R.B. Olsen and O.V. Pictet,
Modelling short-term volatility with GARCH and HARCH models,
in ``Nonlinear Modelling of High Frequency Financial Time Series,'' by C.
Dunis and B. Zhou (John Wiley \& Sons, 1998).

Z. Ding, C.W.J. Granger and R.F. Engle,
A long memory property of stock market returns and a new model,
J. Empirical Finance 1 (1), 83-106 (1993).

R. Ferguson and Y. Simaan, Portfolio composition and the investment
horizon revisited, Journal of Portfolio Management 22, 62-67 (1996).

A. Fisher, L. Calvet and B.B. Mandelbrot, Multifractality of the
deutschmark/us dollar exchange rate, Cowles Foundation Discussion Paper, 1997.

U. Frisch,
``Turbulence: the legacy of A.N. Kolmogorov'' (Cambridge, New York: Cambridge
University Press, 1995).

S. Ghashghaie, W. Breymann, J. Peinke, P. Talkner and Y. Dodge,
Turbulent cascades in foreign exchange markets, Nature 381, 767-770 (1996).

N. Gressis, G.C. Philippatos {\it et al.},
Multiperiod portfolio analysis and the inefficiency of the market portfolio,
Journal of Finance 31, 1115-1126 (1976).

D. Gunthorpe and H. Levy, Portfolio composition and the investment horizon,
Financial Analysts Journal 50, 51-56 (1994).

P. Holgate, The lognormal characteristic function, Commun. Statist.-Theory
Meth. 18, 4539-4548 (1989).

K. Ivanova and M. Ausloos,
Low q-moment multifractal analysis of Gold price, Dow Jones Industrial
Average and BGL-USD exchange rate, Eur. Phys. J. B 8, 665-669 (1999);
erratum Eur. Phys. J. B 12 613 (1999).

Y. Kroll, H. Levy and H.M. Markowitz, Mean-Variance versus direct
utility maximization, The Journal of Finance, vol. XXXIX, N 1, 47-61 (1984).

H. Levy and H.M. Markowitz, Approximating expected utility by a function
of mean and variance, American Economic Review 69, 308-317 (1979).

B.B. Mandelbrot, Fractals and scaling in finance : discontinuity,
concentration, risk,
Selecta volume E (with foreword by R.E. Gomory, and
contributions by P.H. Cootner {\it et al.}), New York : Springer, 1997.

B.B. Mandelbrot, A multifractal walk down Wall Street,
Scientific American 280 N2:70-73 (1999 FEB).

R.N. Mantegna and H.E. Stanley,
An introduction to econophysics: correlations and complexity in finance
(Cambridge, U.K.; New York: Cambridge University Press, 2000).

H. Markovitz, Portfolio selection : Efficient diversification of
investments (John Wiley and Sons, New York, 1959).

J.F. Marshall, The role of the investment horizon in optimal portfolio
sequencing (an intuitive demonstration in discrete time), Financial Review
29, 557-576 (1994).

L. Martellini, Efficient option replication in the presence of transaction
costs, in press in Review of Derivatives Research (2000).

R. C. Merton, Continuous-time finance (Blackwell, Cambridge,1990).

U.A. M\"{u}ller, M.M. Dacorogna, R. DavÈ, R.B. Olsen, O.V. Pictet and J.E. von
Weizs"cker,
Volatilities of different time resolutions - Analyzing the dynamics of market
components, Journal of Empirical Finance 4, No. 2-3, 213-240 (1997).

J.-F. Muzy, J. Delour and E. Bacry, Modelling fluctuations of financial
time series: from cascade process to stochastic volatility model,
submitted to Eur. Phys. J. B,
preprint cond-mat/0005400 (2000).

J.K. Ord, Kendall's advanced theory of statistics. 6th ed.,
Edward Arnold, London and Halsted Press, New York (1994).

M. Pasquini and M. Serva, Clustering of volatility as a multiscale phenomenon,
Eur. Phys. J. B 16, 195-201 (2000).

F. Schmitt, D. Schertzer and S. Lovejoy, Multifractal analysis of foreign
exchange data, preprint. Mc Gill University, Montreal (1999).

D. Sornette,
Large deviations and portfolio optimization, Physica A 256, 251-283 (1998).

D. Sornette, Critical Phenomena in Natural Sciences
(Chaos, Fractals, Self-organization and Disorder: Concepts and Tools)
432 pp., 87 figs., 4 tabs  (Springer Series in Synergetics)
Date of publication: August 2000.

D. Sornette,  P. Simonetti and J.V. Andersen,
$\phi^q$-field theory for Portfolio optimization: ``fat tails'' and
non-linear correlations, Physics Report 335 (2), 19-92 (2000).

G.V.G. Stevens, On Tobins multiperiod portfolio theorem.
Review of Economic Studies V39, 461-468 (1972).

G.Y.N. Tang,
Effect of investment horizon on international portfolio diversification,
International Journal of Management 12, 240-246 (1995).

J. Tobin, The theory of portfolio selection, in ``The Theory of Interest
Rates'', F. Hahn and F. Breechling (eds), London: Macmillan (1965).

N. Vandewalle and M. Ausloos,
 Multi-affine analysis of typical currency exchange rates,
 Eur. Phys. J. B 4, 257-261 (1998).

J. von-Neumann and O. Morgenstern, Theory of games and economic behavior
(Princeton, Princeton University Press, 1944).

G. Zumbach, M. Dacorogna, J. Olsen and R. Olsen, Shock of the new,
RISK, pp. 110-114 (March 2000).

\pagebreak

\pagebreak

\pagebreak

\pagebreak

\pagebreak

\pagebreak

\begin{figure}
\label{fig8}
\begin{center}
\epsfig{file=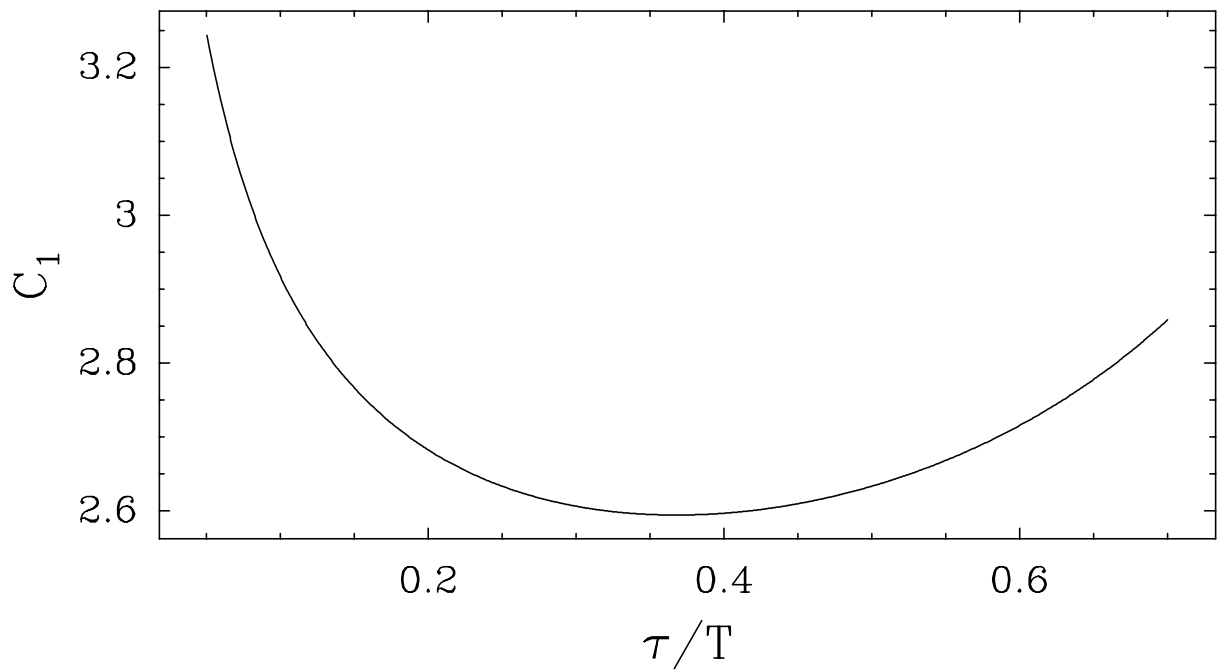, width=16cm}
\caption{\protect\label{figcap8} Return quantified by
$C_1^P(T_h,\tau)$ as a function of the reduced time scale $\tau/T$ for
a given value of $C_4^P=2$. The data are the same as in Fig. 7.
One can note that for a given risk, return is better at very small $\tau$'s
than at very large ones. Moreover, there is a worst horizon for which return is
minimum. 
}
\end{center}
\end{figure}

\end{document}